\documentclass{article}

\usepackage[utf8]{inputenc} 
\usepackage[T1]{fontenc}    
\usepackage{microtype}
\usepackage{times}
\usepackage{graphicx}
\usepackage{amsmath,amssymb,mathbbol}
\usepackage{algorithmic}
\usepackage{acronym}
\usepackage{enumitem}
\usepackage{float}
\usepackage{subcaption}
\usepackage{balance}
\usepackage{xspace}
\usepackage{setspace}
\usepackage[skip=3pt,font=small]{subcaption}
\usepackage[skip=3pt,font=small]{caption}
\usepackage[dvipsnames]{xcolor}
\usepackage{booktabs,tabularx,colortbl,multirow,array,makecell}
\usepackage{pifont}
\usepackage{hyperref}
\usepackage{xr-hyper}
\usepackage[capitalise]{cleveref}
\usepackage{siunitx} 
\usepackage{soul} 
\usepackage{natbib}
\usepackage{svg}
\usepackage{subcaption}
\usepackage{xcolor}

\usepackage{array}
\newcolumntype{P}[1]{>{\centering\arraybackslash}p{#1}}
\newcolumntype{M}[1]{>{\centering\arraybackslash}m{#1}}

\DeclareCaptionLabelFormat{myformat}{\tablename\ #2}
\makeatletter
\renewcommand\p@subtable{} 
\makeatletter
\newcommand{\cmark}{\ding{51}}%
\usepackage[accepted]{icml2021}
\icmltitlerunning{Alexa Arena: A User-Centric Interactive Platform for Embodied AI}

\begin{document}

\twocolumn[
\icmltitle{Alexa Arena: A User-Centric Interactive Platform for Embodied AI}



\icmlsetsymbol{equal}{*}

\begin{icmlauthorlist}
\icmlauthor{Qiaozi Gao}{equal,az}
\icmlauthor{Govind Thattai}{equal,az}
\icmlauthor{Suhaila Shakiah}{equal,az}
\icmlauthor{Xiaofeng Gao}{equal,az}
\icmlauthor{Shreyas Pansare}{az}
\icmlauthor{Vasu Sharma}{ex-alexa}
\icmlauthor{Gaurav Sukhatme}{az}
\icmlauthor{Hangjie Shi}{az}
\icmlauthor{Bofei Yang}{ex-alexa}
\icmlauthor{Desheng Zheng}{az}
\icmlauthor{Lucy Hu}{az}
\icmlauthor{Karthika Arumugam}{az}
\icmlauthor{Shui Hu}{ex-alexa}
\icmlauthor{Matthew Wen}{az}
\icmlauthor{Dinakar Guthy}{az}
\icmlauthor{Cadence Chung}{az}
\icmlauthor{Rohan Khanna}{az}
\icmlauthor{Osman Ipek}{az}
\icmlauthor{Leslie Ball}{az}
\icmlauthor{Kate Bland}{ex-alexa}
\icmlauthor{Heather Rocker}{az}
\icmlauthor{Yadunandana Rao}{az}
\icmlauthor{Michael Johnston}{az}
\icmlauthor{Reza Ghanadan}{az}
\icmlauthor{Arindam Mandal}{az}
\icmlauthor{Dilek Hakkani Tur}{az}
\icmlauthor{Prem Natarajan}{az}

\end{icmlauthorlist}

\icmlcorrespondingauthor{Qiaozi Gao}{qzgao@amazon.com}
\icmlcorrespondingauthor{Govind Thattai}{thattg@amazon.com}
\icmlcorrespondingauthor{Xiaofeng Gao}{gxiaofen@amazon.com}
\icmlcorrespondingauthor{Suhaila Shakiah}{ssshakia@amazon.com}

\icmlaffiliation{az}{Amazon Alexa AI}
\icmlaffiliation{ex-alexa}{Work done while at Amazon Alexa AI}


\vskip 0.3in
]



\printAffiliationsAndNotice{\icmlEqualContribution} 

\begin{figure*}[t]
    \centering
    \includegraphics[width=\textwidth]{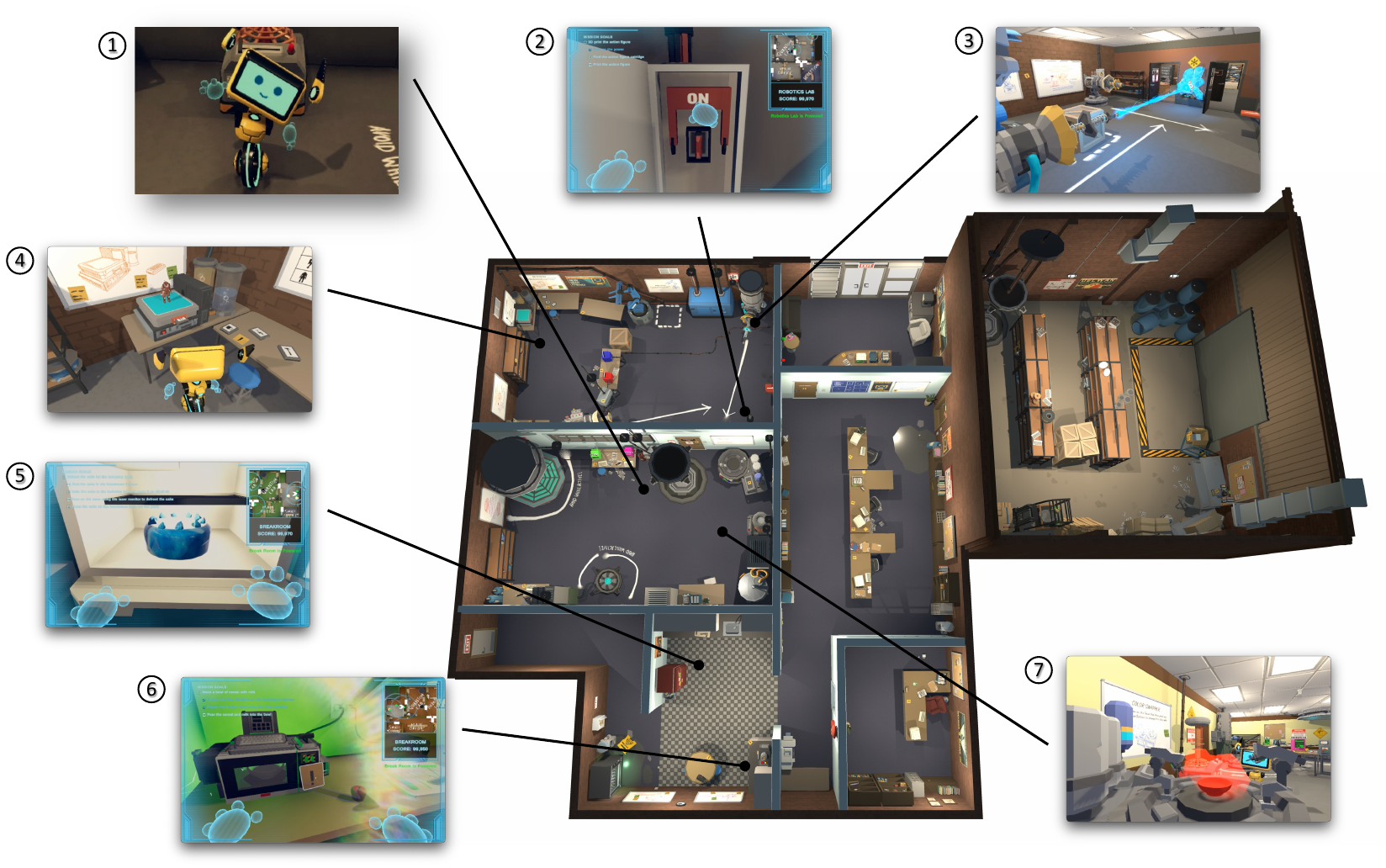}
    \caption{As an EAI platform, Alexa Arena has a variety of object categories, including a set of fantastical objects as engagement enhancers, such as the freeze ray (3), the time machine (6) and the color changer (7). As a result, the agent (1) can make use of different tools to change the state of the object. For example, the agent can use either the freeze ray (3) or the fridge (5) to cool an object. Arena also has more interaction actions and state changes compared to existing platforms. For instance, the agent can use the time machine to repair broken objects (6), or use the color changer to change the color of objects (7).}
    \label{fig:teaser}
\end{figure*}
\begin{abstract}

\textbf{We introduce Alexa Arena, a user-centric simulation platform for Embodied AI (EAI) research. Alexa Arena provides a variety of multi-room layouts and interactable objects, for the creation of human-robot interaction (HRI) missions. With user-friendly graphics and control mechanisms, Alexa Arena supports the development of gamified robotic tasks readily accessible to general human users, thus opening a new venue for high-efficiency HRI data collection and EAI system evaluation. Along with the platform, we introduce a dialog-enabled instruction-following benchmark and provide baseline results for it. We make Alexa Arena\footnote{\href{https://github.com/amazon-science/alexa-arena}{https://github.com/amazon-science/alexa-arena}} publicly available to facilitate research in building generalizable and assistive embodied agents.}
\end{abstract}

\section{Introduction}
A longstanding goal of AI is to develop autonomous robotic agents that can assist humans in day-to-day activities. For multiple reasons, experiments with robots are often conducted in controlled environments, which limit their variety and scale. To mitigate this problem, several embodied AI simulation platforms have recently been proposed. These platforms support a number of virtual scenes that can be either manually designed, synthetically generated, or captured from real scenes. An embodied agent can freely navigate and interact with objects in these scenes to complete tasks. However, the current EAI platforms suffer from a set of limitations that curtail the ability to build generalizable assistive AI agents. 

\textbf{Facilitating HRI data collection by gamification.}
A persistent challenge in HRI is collecting human interaction data including natural language instructions along with visual content and actions. Currently available EAI platforms, however, are not designed for humans interacting with the agent. As a result, the data collection process is often expensive and time consuming \cite{mandlekar2018roboturk, srivastava2022behavior}. Games have been historically utilized to encourage wide-spread user participation and engagement with services \cite{usinggames, ANGELESQUIROGA20151, Ram2007ArtificialIF}. To ease the data collection effort, we introduce gamification to EAI, which is achieved by presenting tasks or user interactions in the form of games, by introducing scoring mechanisms, achievements and streaks to stimulate and engage users, thus promoting and more importantly, retaining their participation. 

\textbf{Reasoning based on visual observations.} 
Tasks in current EAI platforms have limited in-class variability: each type of task requires the agent to make use of the same object repeatedly. As a result, agents may resort to over-engineering (over-fitting to specific and simplistic tasks) and memorization to converge to trivial and non-generalizable solutions or use predefined task decomposition mappings \cite{min2022film}. One way to mitigate this problem is to design missions that can be completed in multiple ways by interacting with different objects in the environment with compositional and causally interconnected state changes. To complete the mission efficiently, the agent is then required to understand its current state from visual observations and choose the subsequent approach accordingly. 

To this end, we propose Alexa Arena, a user-centric EAI platform. For better user experience, the platform design has commonalities with games, with features like task-guiding UI elements and engaging visual effects. The platform boasts many variations of objects and state transitions, on top of which a variety of missions are designed (\Cref{fig:teaser}). We show one example use case of Arena on dialog-guided task completion, where the embodied agent can communicate with the user through natural language to finish an indoor mission. To help develop models for this task, we propose a dataset which includes an expert demonstration for each mission and the corresponding human-annotated language instructions and dialogues. To evaluate the dialog-guided agent in real-time, we also release a web-based user interface (UI) as a proof of concept, where a user can communicate with the agent in text and receive visual observations of the agent from Arena. 

This paper makes the following contributions: 
\begin{enumerate}[leftmargin=*]
\item Alexa Arena as a new user-centric EAI platform which focuses on building generalizable agents that can assist humans through reasoning and procedural learning.
\item A dialog-guided task completion benchmark created from Alexa Arena, including 3K unique tasks and 46K human annotated instructions and dialogues. We also present results from baseline models on the benchmark.  
\item A web-based user interface allowing real-time user communications and interactions with an EAI agent in Arena. 
\end{enumerate}

\begin{table*}[t]
    \centering
    \begin{tabular}{lccccccccc}
        \hline
         &  & Object  & & Object & 3D & Real &  Human & Game
         \\ Platforms & Scenes & Types & Multi-room & States & Env & Scenes & Control & Ready \\ \hline
        Malmo \cite{johnson2016malmo}  & 17*   &  1000+  &    &   \cmark  & \cmark &  &  MK   & \cmark  \\ 
        Overcooked \cite{carroll2019utility}  & 6*   &  5  &   &   \cmark  &   &   &  MK   & \cmark  \\ 
        Deepmind Lab \cite{beattie2016deepmind} &  28*   & -   &    &   & \cmark &   &  MK  & \cmark \\ 
        AI2-Thor \cite{kolve2017ai2}  & 120    & 125+   &    &   \cmark  & \cmark & \cmark  & MK,VR \\ 
        iGibson \cite{xia2018gibson}   & 15     & 390+  &  \cmark &  \cmark  & \cmark & \cmark  & MK,VR \\ 
        Habitat 1.0 \cite{savva2019habitat} & 1000   & -  &   \cmark &  & \cmark & \cmark  &  MK    \\
        Habitat 2.0 \cite{szot2021habitat} & 105 & 90+  & \cmark & \cmark & \cmark & \cmark  &  MK\\
        VirtualHome \cite{puig2018virtualhome} & 7    & 170+   &   \cmark &  \cmark & \cmark & \cmark & NL \\
        TDW \cite{gan2020threedworld} & 15 & 200+ & & \cmark & \cmark & \cmark & VR \\ 
        VRKitchen \cite{gao2019vrkitchen} & 16 & 200+ & & \cmark & \cmark & \cmark  &  VR \\ \hline 
        \textbf{Alexa Arena}     & 10 & 335+ & \cmark  & \cmark  &  \cmark  & \cmark  &  MK,NL  & \cmark \\ \hline
    \end{tabular}
    \caption{Comparison with other platforms. Scenes and objects: the number of scenes and interactable object types the platform provide.  Multi-room: multi-room scenes. Some platforms marked by * are customizable and new scenes can be added. States: agent can change object states via interactions. 3D env: whether the platform supports 3D virtual environment. Real scenes: whether the scenes in the virtual environment are realistic. Human Control: the interface for the human to control the agent. MK: mouse and keyboard. NL: natural language. VR: virtual reality. Game ready: whether the platform has game-ready features to enhance user experience, including environmental and interaction sounds, action animations, smooth navigation, fantastical objects, hazards, mini-map, scoring mechanism and hints.}
    \label{tab:comparison}
\end{table*}

\section{Related Work} 

\noindent \textbf{Embodied AI platforms.} 
In recent years, many platforms have been proposed for embodied agents to perform household activities in indoor virtual environments \cite{ puig2018virtualhome, gan2020threedworld, xiang2020sapien}. Most simulators are designed with performance and realism as the top priorities \cite{szot2021habitat}. Although some simulators have built-in infrastructure for human data collection \cite{gao2019vrkitchen, xia2018gibson}, the user interactions are generally not the focus of their designs.   On the contrary, Alexa Arena is a user-centric embodied AI platform with features designed specifically to enhance user experience: 1) the platform includes engage-enhancers, such as fantastical objects and a scoring system for each mission; 2) for better visual effects, actions in Arena are animated in a continuous played-out fashion for both manipulation and navigation actions; 3) the platform features a unique user interface with several elements to provide users with better task guidance, including minimap, sub-task hints and sticky notes, all of which work in an integrated fashion to improve user-experience. There are also several EAI platforms that are inspired by video games \cite{johnson2016malmo, beattie2016deepmind, tian2017elf, carroll2019utility, gao2020joint, yuan2022situ}, but the scenes are often unrealistic and the agent visual perceptions are simplified. See \Cref{tab:comparison} for a comparison between Alexa Arena and other embodied AI platforms.


\noindent \textbf{Language-guided navigation and task completion.} Most existing work for learning language-guided embodied agents focuses on navigation tasks \cite{anderson2018vision, nguyen2019help, chi2020just, roman2020rmm}. For increased task complexity, \cite{misra2018mapping, shridhar2020alfred} enable the agent to follow natural language instructions and complete household activities, where both navigation and object manipulation actions are required. Most recently, \cite{padmakumar2021teach, gao2022dialfred} propose new datasets and benchmarks for training and evaluating task-oriented embodied agents that can engage in dialogue. However, based on the AI2-Thor simulator, both works use offline settings where the dialogues are either pre-collected from humans or generated with templates. In comparison, the Arena platform enables embodied agents to communicate with human users in a real-time interactive fashion to finish the task. 

\noindent \textbf{Task planning with large language models.}
Recently, there is a growing trend of using large language models (LLMs) for assisting robot task planning in learning novel activities or completing complex tasks \cite{ahn2022can, wang2023describe}. LLMs have been shown to be good at providing high-level semantic knowledge about the physical world and common human activities \cite{huang2022language, singh2022progprompt}. When combined with the sensory perception of the embodied agent, such knowledge can often substantially improve the agent's capability of solving complex tasks in unseen scenarios \cite{inoue2022prompter, huang2022inner}. The Arena platform provides a good testbed for this line of work. With numerous object types, properties and state changes, along with various environmental causal events, Arena supports the creation of robot tasks that require adequate reasoning capability on common world knowledge.

\section{The Arena Platform} 

In this section, we describe the Alexa Arena Platform. 
We briefly describe the overall framework, attributes of the simulator and tools that we are releasing as a part of the platform to bootstrap development with Arena. 

\begin{figure}[t]
    \centering
    \includegraphics[width=0.45\textwidth]{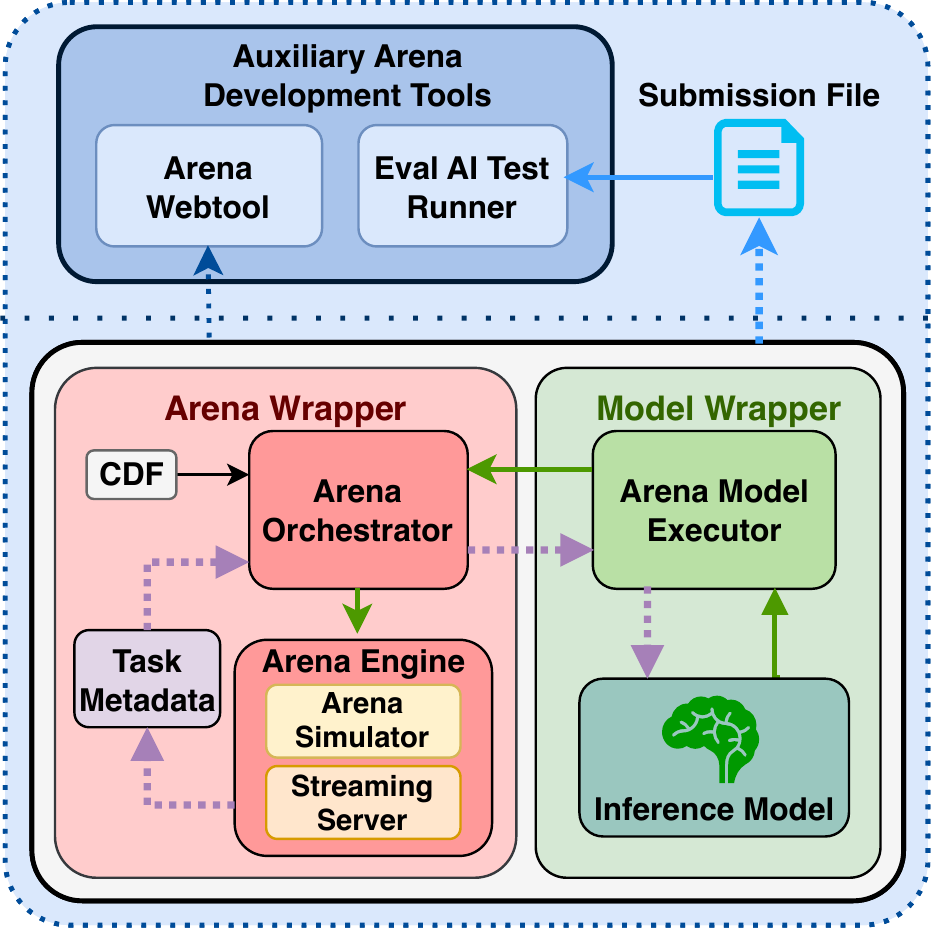}
    \caption{Overview of Alexa Arena Platform Architecture. The core modules are the \textit{arena wrapper} and \textit{model wrapper}. The arena engine contains the \textit{simulator} and the \textit{streaming server}. The \textit{metadata} from the Arena Engine are sent to the inference model, which generates actions to be executed in the simulator. We have also built infrastructure to generate \textit{results submission files} for EvalAI challenges.}
    \label{fig:arena_archt}
\end{figure}

\subsection{Architecture}
The architecture of the Alexa Arena framework is illustrated in Figure~\ref{fig:arena_archt}. The core components of the platform are the \textit{arena wrapper} and \textit{model wrapper}. The \textit{arena wrapper} houses the \textit{arena orchestrator}, which takes as input the configuration file (described in Section \ref{cdf}), and communicates with the arena engine, which contains the the \textit{arena simulator} and the \textit{streaming server}. The \textit{streaming server} in the arena engine streams the robot's first party camera view to human users. The \textit{model wrapper} contains the \textit{inference model} and the \textit{model executor} that interfaces between the model and the orchestrator. The \textit{model wrapper} can be modified to develop other models and execution strategies. In addition, we also provide a web interface (the \textit{webtool}) for interacting with a robot model in a chatbot like interface, while also watching the game play out in streaming on a web page (\Cref{fig:web_tool}). 

\subsection{Challenge Definition Format}
\label{cdf} 
To define the mission configurations, we introduce a schema called Challenge Definition Format (CDF). A CDF file is a json-format file that is readable by Arena. It contains the necessary information to configure, initialize and run a game mission in Arena. These include:

\noindent \textbf{1. Game initial state:} game scene setting, robot initial location, object initial location and initial states (such as cabinet door being open or closed).

\noindent \textbf{2. Game goal condition:} robot and object states that need to be satisfied. Once the goal condition is met, the game mission is considered completed.

\noindent \textbf{3. Game-related text data:} textual data that help guide the user to finish the game, including game mission description text, sub-goal descriptions, system prompts during game play, etc.

The motivation for designing and utilizing the CDF is to allow programmatic generation of large numbers of game missions, which is essential for data collection. Since manually defined game missions are not scalable for generating large-scale training data for ML models, with the easily configurable CDF files, we can generate thousands of game missions by permuting scenes, robot location \& states, object types, object location \& states, and mission goal states. The other advantage is that since the CDFs are human readable, we can specify new game missions flexibly without writing code. 

\subsection{Environment Metadata} 
The environment metadata in Arena contains robot camera view images, agent state, objects state, scene metadata, goal condition status, and previous action execution result. To facilitate different task and modeling settings, the robot camera view images support different modalities, including RGB image, depth image, instance segmentation image and normals image. The metadata response from the arena simulator is updated after each action execution.

\subsection{Objects Properties and States}
There are 336 unique objects in Arena. Each object has a set of properties (i.e. affordances), which specify if a certain type of robot-object interaction is possible. For example,  the agent can toggle the \textit{3-D printer} since it has an object property \textit{toggleable}. At the same time, the agent cannot pick it up since it does not have the \textit{pickupable} property. In total, there are 14 object properties, including \textit{pickupable, openable, breakable, receptacle, toggleable, powerable, dirtyable, heatable, eatable, chillable, fillable, cookable, decor} and \textit{infectable}. Each object property has a corresponding action and object state upon acting. For example, \textit{break} is the corresponding action for \textit{breakable}, and \textit{broken} is the corresponding state after the action has been performed. The states of objects will change as a consequence of the action, given that the pre-conditions are met. For example, if powered on (\cref{fig:teaser} (2)), the \textit{3-D printer} can be used to make toys (\cref{fig:teaser} (4)).

To further mimic the complexities of real-world tasks and execution strategies, along with numerous object types and properties, Arena also provides a variety of constraints defined by commonsense causal events. For example, the power system enables a series of causal constraints on using electric appliances, e.g., the fuse box needs to be reset before lights being turned on, the microwave needs to be plugged in to the outlet before it can be powered on. Agents in Arena also have multiple ways to complete a task, e.g., To heat something, the agent can make use of either the \textit{Laser} or the \textit{Microwave}. As a result, the agent needs to reason based on its current state for efficient task completion. As an example, consider the task of heating a bowl and delivering it to a receptacle in the \textit{Quantum Lab}, and the agent is currently in the \textit{Quantum Lab}. If the agent's execution path to heat the bowl using the \textit{Laser} in the \textit{Quantum Lab} is physically obstructed, it is more efficient for the agent to go to the \textit{Breakroom}, heat the bowl quicker using the \textit{Microwave} and return to deliver the bowl in the \textit{Quantum Lab}. While this may seem inefficient at the outset (to go to another room as opposed to completing the mission with an object in the current room), experience-based long-term reasoning could confer the agent with the skills to take efficient steps to complete the mission. With such flexible task completion enabled by causal constraints, object variations and state transitions, Arena can be used as a test bed for situated commonsense reasoning and high level task planning.

\subsection{Robot Action Space}
In general, Arena supports two kinds of actions: 1) user interaction actions for communicating with the user both textually and visually, and 2) robot physical actions to interact with the simulation environment.  There are two types of robot physical actions - navigation actions and object interaction actions. For better user experience, all the navigation and interaction actions are animated in a continuous fashion and environmental sounds are played during the animation.

\begin{figure*}[t]
    \centering
    \includegraphics[width=\textwidth]{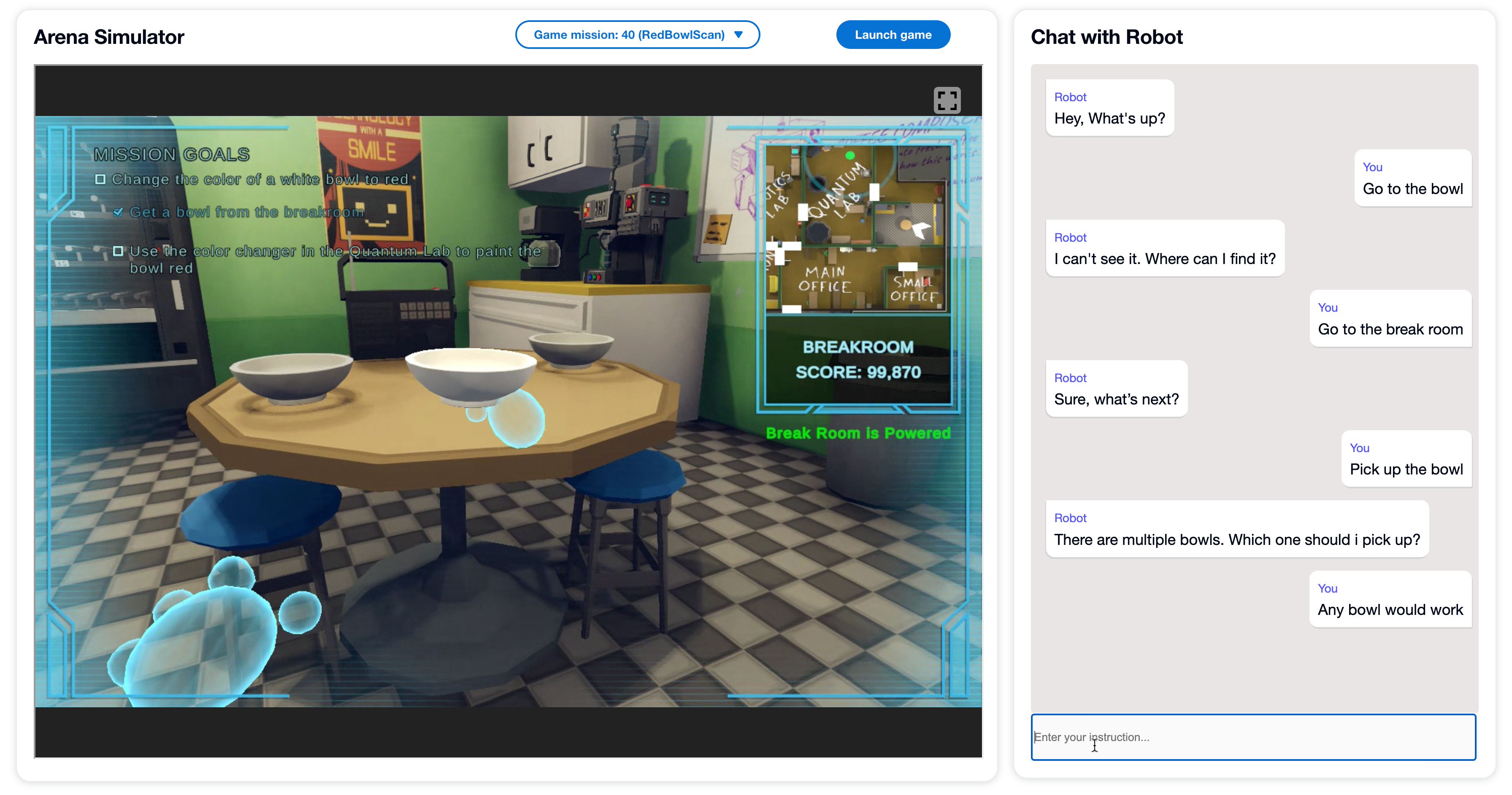}
    \vspace{-0.5cm}
    \caption{The web-based user interface. The user can use the chat box to communicate with the agent in Arena for task completion. Meanwhile, video from Arena is streamed in real time to show the progress. The minimap is displayed on the top right corner of the game UI, showing the room layout and the robot location and orientation. The mission goal and subgoals are displayed to the users in text on the top left corner of the UI. }
    \label{fig:web_tool}
\end{figure*}

\textbf{User interaction.} To communicate with the user, the robot can initiate a dialog, the contents of which are displayed on the user interface (\Cref{fig:web_tool}). The robot can also \textit{highlight} objects for real-time visual confirmation and language instruction disambiguation.

\textbf{Navigation.} The goal of Arena is to focus specifically on compositional task learning and reasoning, instead of indoor navigation. To this end, we simplify the navigation in Arena by enabling the robot to directly navigate to a viewpoint in a room by specifying the viewpoint name or the room name, or to an object by specifying the object mask. Meanwhile, if preferred, the robot can also perform step-by-step navigation by a combination of local primitive actions like \textit{MoveFoward}, \textit{MoveBackward}, and \textit{Rotate} actions which can take granular inputs as arguments. To help the robot perceive its immediate surroundings and navigate to objects in its vicinity, the platform also supports a special \textit{lookAround} action that enables the robot to get panoramic images.

\textbf{Object interaction.} Arena supports 11 actions for object interaction, including \textit{Examine, Pickup, Place, Open, Close, Break, Pour, Toggle, Fill, Scan} and \textit{Clean}. Each action is associated with a set of objects in a specific state in which the objects can afford that action to be performed upon them. E.g., \textit{toggling} can be performed on a \textit{Time Machine} when it is in a \textit{closed} state. Interaction actions are accompanied by a change in the associated object's state, which gets updated in the environment metadata. 

\begin{figure*}[t]
    \centering
    \includegraphics[width=\textwidth]{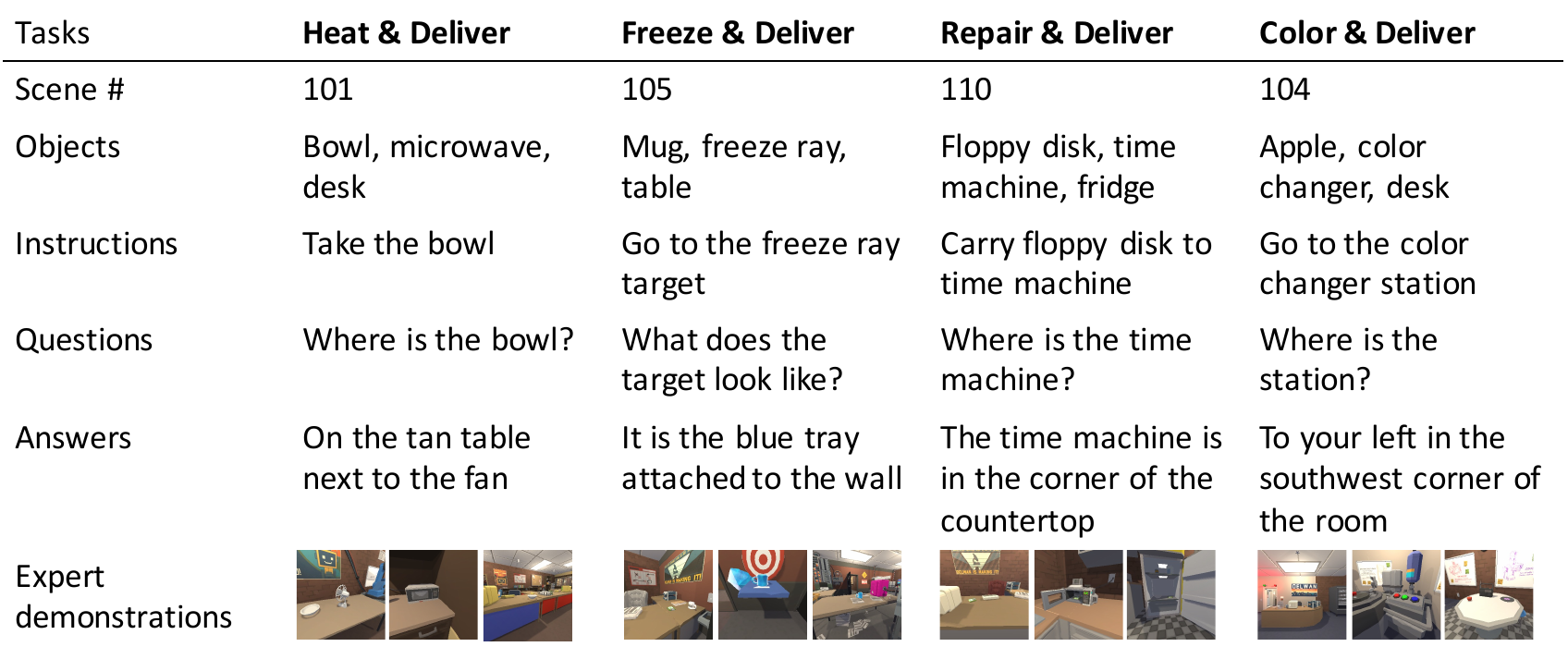}
    \caption{Examples in our trajectory dataset. Each data session corresponds to one language annotation on a mission that needs to be completed in a specific scene. Each language annotation includes a sequence of instructions guiding the robot to complete the task, as well as questions and answers for clarification. Expert demonstrations are provided for developing the model. }
    \label{fig:tasks_examples}
\end{figure*}

\subsection{Enhanced User Experience}
Improved user experience is one of the hallmarks of the Arena platform, aiming to bootstrap the data collection and EAI evaluation process with human users in the loop. Therefore, we incorporate into Arena elements that enhance application usability and improve engagement quality. 
\begin{enumerate}[leftmargin=*]

\item We provide a \textit{Sticky Note} object that can be populated with textual hints and guidelines for the mission. \textit{Examine} is a special action designed for the user to interact with the \textit{Sticky Note} object to read the hints, if needed, for completing the mission goals. The hints mechanism can be used as an auxiliary aid to signal the user interacting with the agent for purposes of data collection or educating the user about the environment or task goals (these are not considered to be a part of the agent algorithm or the dataset as a result). This can be configured using the CDF.
\item The user interface also allows displaying the goal and sub-goal texts on screen which remind the users of the subgoals that need to be completed in order to complete the mission. This functionality also displays a check mark for a goal if it is completed. This is different from the \textit{Sticky Note}, which is designed to guide users to finish the mission. 
\item We also embed a minimap in the user interface which displays the real-time position and orientation of the robot in the layout, along with the room names and a top view of the floor plan (\Cref{fig:web_tool}). It also features strobing signals to capture human attention and guide the users to the location of the \textit{Sticky Notes}. 
\item The platform showcases high quality graphics, allowing streaming with continuous animations, i.e., actions like opening and closing an object, or navigating from one point to another are animated and seamlessly played out in real-time to the user. In addition, to induce a more engaging and vivid user experience, the environment also animates certain object states like a sparking power cable, steam coming off a hot coffee, or a laser beaming through the air. These animations are visible for objects being interacted with, as well as for objects in the background that are within the agent's field of view. This makes the scenes more natural, user-friendly and engaging during mission play.
\item To enhance and retain user participation, we gamify the missions on Arena. In addition to the incorporation of fun and fantastical objects like \textit{Time Machine} and \textit{Color Changer}, we also have a scoring mechanism which is displayed below the minimap (\Cref{fig:web_tool}). This is designed to be a count down system that reduces in value as time passes by. This can be used to motivate users to achieve higher scores and also to evaluate the agent. 
\end{enumerate}

\section{A Dialog-guided Task Completion Benchmark} 
\label{challenge}
We now present the Dialog-guided Task Completion benchmark. The benchmark is designed to evaluate dialog-guided agents for indoor object interaction tasks. To support model development, we release a hybrid dataset where ground-truth robot action trajectories are paired with human annotated language. We also set up a challenge\footnote{\href{https://eval.ai/web/challenges/challenge-page/1903/}{https://eval.ai/web/challenges/challenge-page/1903/}} on EvalAI \cite{yadav2019evalai} and encourage researchers to participate. 

\subsection{Task settings} 

Tasks in the challenge require the agent to follow natural language instructions from the user, and perform navigation and object manipulation actions in a virtual environment. Each task instance is specified by the start states and the goal completion states of a scene, and a sequence of human instructions. The goal of the agent is to interact with the environment through a series of actions by following the human instructions to achieve the goal states. There are 12 unique tasks types in the challenge, including \textit{pickup\&deliver}, \textit{heat\&deliver}, \textit{freeze\&deliver}, \textit{repair\&deliver}, \textit{fill\&deliver}, \textit{color\&deliver}, \textit{clean\&deliver}, \textit{pourContainer}, \textit{breakObject}, \textit{insertInDevice}, \textit{toggleDevice} and \textit{scanObject}. Examples of data points in a few task types are illustrated in \Cref{fig:tasks_examples}.

\begin{figure}[t]
    \centering
    \includegraphics[width=0.50\textwidth]{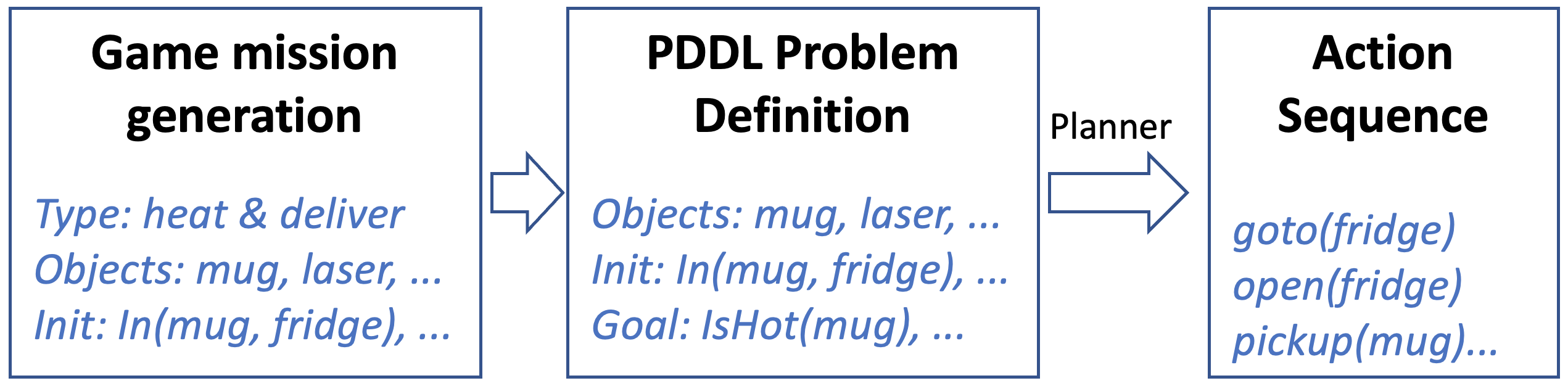}
    \caption{Process of generating game missions and expert demonstrations.}
    \label{fig:data_generation}
\end{figure}

\subsection{Trajectory Dataset}

\begin{table}[t]
\centering
\begin{tabular}{lrrrrr}
           & \# scene & \# task & \# session & \# instr & \# dialog \\ \hline
Train      & 6  & 2661 & 7983  & 40443 & 40105     \\
Valid      & 6 & 383 & 1149 & 6120 & 6128   \\  \hline
\textbf{Total}      & 6 & 3044 & 9132 & 46563 & 46233  \\
\hline
\end{tabular}
\caption{Trajectory Dataset Breakdown. \# session represent the number of data sessions. Each data session corresponds to one language annotation on a given task. \# instr stands for the total number of instructions in each data split. \# dialog is for the total number of question and answers. }
\label{tab:traj_breakdown}
\end{table}

To enable training and evaluation of the embodied agent, we collect a dataset containing expert action trajectories, human language instructions and questions and answers for  missions. We split the demonstrations and human language annotations into training and validation folds. There are 2661 tasks in training and 383 in validation, each task are annotated by 3 annotators. Each human annotation is considered one episode, which gives a total of 7983 training episodes and 1149 validation episodes (\Cref{tab:traj_breakdown}). 

\subsubsection{Expert demonstrations}


The game mission and expert demonstration generation process is illustrated in Figure~\ref{fig:data_generation}. In step one, game missions are programmatically generated via sampling from initial environment states (e.g., game scene, robot initial location, task-related objects and their states) and mission goals. Once the game missions are defined, the next step is to generate expert demonstrations using a planner. Along with the game definitions, we also generate PDDL (Planning Domain Definition Language) planning problem definitions for each game mission. A symbolic planner is used to solve the planning problem and the output action sequence is collected as the expert demonstration. One thing to note is that the planner has access to the game metadata, which is not available to the agents during inference time. For tasks that can be completed in different ways (e.g., by using different tools), we pick one unique way for each task to generate expert demonstration. For example, for freeze and deliver, in some missions the fridge is used to cool the object, while in other cases the freeze ray is used. 

\subsubsection{Human Language Annotation} 

To collect natural language dialogue for the missions, we design a two-stage data annotation process on Amazon Mechanical Turk (AMT). The first stage is for collecting instructions and the second is for questions and answers. In both stages, the annotator watches a video containing an expert demonstration for the mission and provides the annotations in free-form text data or answers to multiple choice questions.

In the first stage, annotators are told to write instructions to tell a “smart robot” how to accomplish a task. During the process, an annotator first watches the video of the ground-truth robot actions, then writes instructions for each highlighted video segment. \cref{fig:ui_instruction} shows the user interface for collecting language instructions. 

After all the instructions are collected, we start the second stage of annotation, where the annotators are asked to raise questions to better complete the task, as if they are controlling the robot to follow the instructions. They also need to subsequently answer their own raised questions. Similar to \cite{gao2022dialfred}, the questions choices in the second stage of annotation are generated using predefined templates. Given an instruction, we extract nouns and insert them into the templates to construct questions. The nouns extracted correspond to query objects that the agent needs more information about to complete the mission. For more details on the human language annotation process, see Appendix \ref{appendix:annotation}.

\subsection{Vision Dataset} 

\begin{figure}[t]
    \centering
    \includegraphics[width=0.5\textwidth]{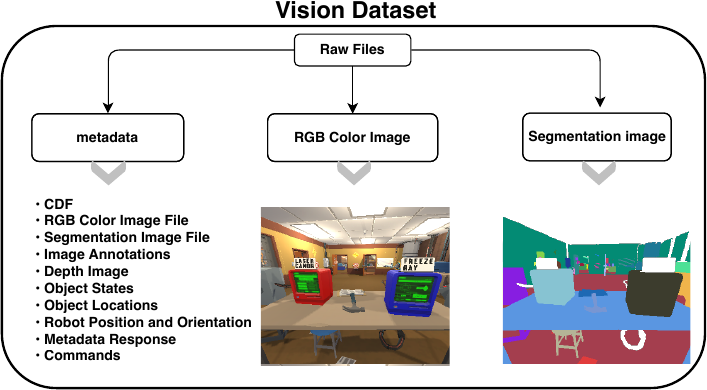}
    \caption{Overview of the vision dataset. The dataset consists of RGB and ground truth segmentation images in png format, along with a json file for each image with the metadata information.}
    \label{fig:vision_dataset}
\end{figure}

Visual perception is integral for EAI agents to navigate and interact within their environments. Visually intelligent robots use visual cues to set navigational targets and build visual representations based on perceived object state transitions. To facilitate visual AI research within and outside the context of EAI, we are releasing a vision dataset based on Arena. The dataset is composed of 600k training images and 60k validation images spanning 336 unique objects from more than 160 semantically grouped object groups. This can be used for large scale computer vision research on Arena, e.g. object detection, object state classification, etc. The overall structure and contents of the dataset are illustrated in \cref{fig:vision_dataset}. The vision dataset is collected by programmatically configuring CDFs and initializing scenes with all objects spawnable in the Arena environment. We then navigate to each of them to capture images from different perspectives and distances.  Detailed vision data generation process is described in Appendix~\ref{appendix:vision_data}.

     


\begin{figure*}[h]
    \centering
    \begin{subfigure}[p]{0.8\textwidth}
        \centering
        \includegraphics[width=\textwidth]{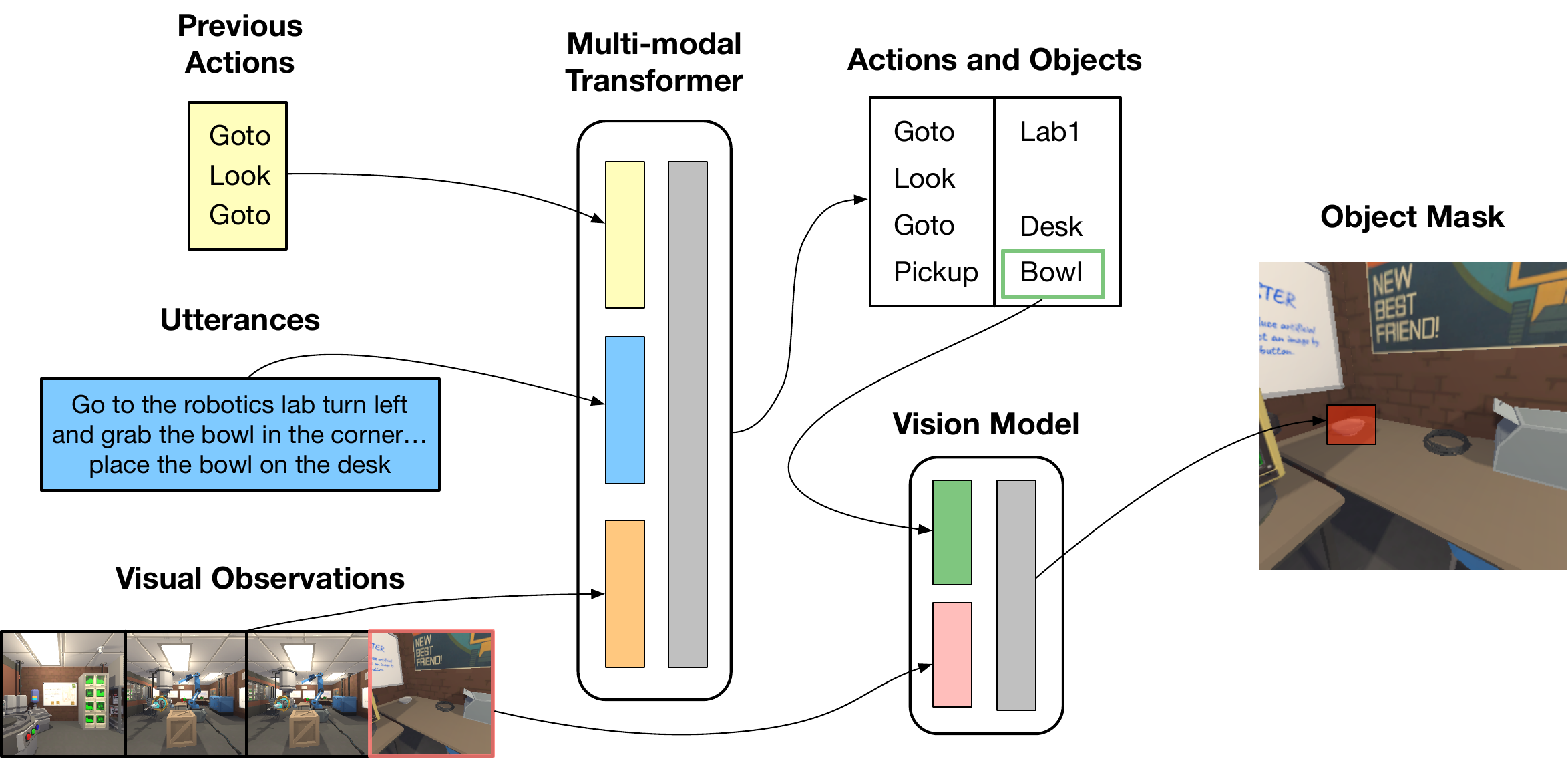}
        \caption{Neural-Symbolic Approach}
        \vspace{.5cm}
        \label{fig:et_model}
    \end{subfigure}
    \begin{subfigure}[p]{0.8\textwidth}
    \centering
    \includegraphics[width=\textwidth]{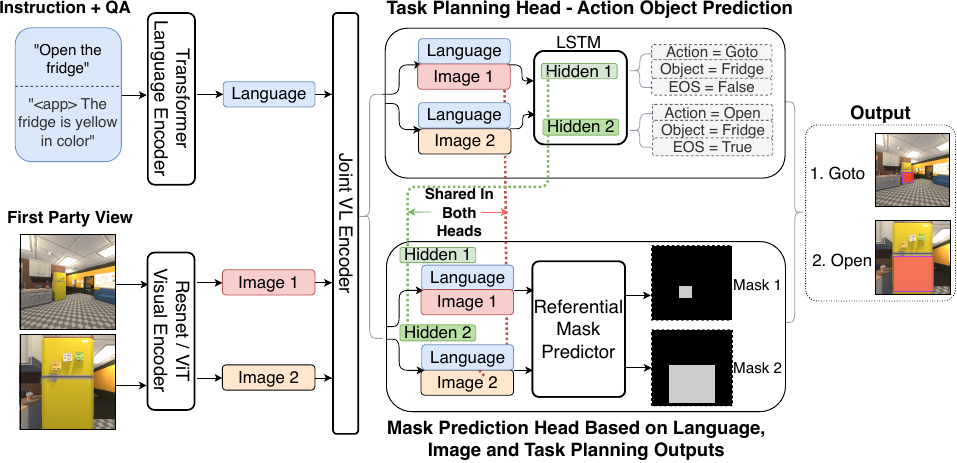}
    \caption{Vision Language Model}
    \label{fig:vl_model}
    \end{subfigure}
    \caption{Model architecture. a) Neural-Symbolic approach which uses unshared modules to predict actions and masks, and heuristic rules for visual grounding. b) A vision language model that does end-to-end action prediction and neural visual grounding.}
    \label{fig:model_architecture}
\end{figure*}

\section{Baseline Models} 

We provide two baseline models trained on the trajectory and the vision datasets, and evaluated on the validation trajectory dataset. The inputs to the models are the first party view of the agent and the natural language instruction. The outputs are a series of action sequences (and associated interactable masks, where necessary) to be executed in the environment to complete the task at hand.



\subsection{A Neural-Symbolic Approach}

Episodic Transformer (ET) is a multi-modal Transformer model for visual-language navigation \cite{pashevich2021episodic}. ET uses the historical visual and language information in the entire episode to capture long-term dependencies between actions. It was originally designed for the Alfred Challenge \cite{shridhar2020alfred}. We applied the Episodic Transformer model on our challenge, with a vision model trained on the Arena dataset for object detection. The architecture is shown in \Cref{fig:et_model}.

\subsubsection{Vision Model}

We use the vision dataset to train a Mask-RCNN image segmentation model \cite{DBLP:journals/corr/HeGDG17}. We process the object classes in the vision dataset to group them into 86 semantic classes including the background class. The model takes as input an RGB image and predicts masks for all object instances (across the 86 object classes) present in the image along with their class labels and confidence scores. The masks and object class predictions are then used in conjunction with the action and object predicted from the multi-modal transformer to give a single mask prediction for the interactable object.

The vision model is trained for 22 epochs with a learning rate of 0.00125 and a weight decay of 0.001 with the SGD optimizer, decaying the learning rate after 15 epochs by a factor of 0.1. We use a global batch size of 16 and trained the model on 2 Tesla V100 GPUs.

\subsubsection{Experimental Setup}

The multi-modal Transformer model uses the previous actions, previous visual observations and the full utterances of the episode to predict the next action type and object class before a \textit{stop} token is predicted as the action. For the visual observations, we first extract features using the backbone of the vision model trained on the vision dataset before sending them into the transformer. During training, after each look around action, the model uses the observations from different directions to predict rotations: the number of rotation depends on which image out of the four panoramic images is used by the following actions in the expert demonstrations. During inference, the vision model detects all the objects from the current image, and uses the predicted object class to generate the mask for the corresponding object. Note that if the vision model detects multiple object instances of the same class, we select the one with the highest likelihood. 

For language utterance, we tried using only the human instructions, as well as the instructions plus the questions and answers as input. We use two tokens to efficiently encode the predefined questions, including (i) a token representing the question type (i.e. \textit{loc} for questions asking about object location, \textit{app} for object appearance,  \textit{dir} for direction and \textit{ref} for reference), and (ii) a token for the target object in location, appearance, referential questions. For example, a question asking about the appearance of a microwave will be encoded as \textit{app microwave}. Since the direction question is asking about the direction and not related to an object, the second token is not necessary and thus omitted for this type of question. For questions that are not of the predefined types, we include the whole question and answers. 

We trained the transformer model for 20 epochs with a batch size of 2. We use the Adam optimizer with a learning rate of $1e^{-4}$ for the first 10 epochs, and $1e^{-5}$ for the remaining epochs.

\subsection{End-to-end Vision language Model}
We experimented with an end-to-end vision language (VL) model to predict the actions and the interactable masks given a natural language instruction and the first party view of the agent. This model uses the disambiguation information present in the natural language instruction to do visual grounding, i.e. it predicts a single mask for the interactable object instance as opposed to using rule-based grounding.

\subsubsection{Experimental Setup}
 The architecture of the model is shown in \Cref{fig:vl_model}. A high level natural language instruction is encoded using a transformer model. For demonstrating the usefulness of the dialog induced questions and answers (QA), we experiment with encoding just the language instruction (without QA), as well as the language instruction, question type and the answers to the associated questions to include contextual and disambiguation information (with QA). We use the same tokens as the ET model to encode the question type.
 The egocentric RGB image is encoded using a ResNet \cite{https://doi.org/10.48550/arxiv.1512.03385} or ViT \cite{dosovitskiy2020image} backbone. We use the ResNet backbone for our experiments. The image and language encoders are initialized with pre-trained weights from CLIP \cite{https://doi.org/10.48550/arxiv.2103.00020}. The visual and language features are then passed through a joint VL encoder that outputs a combined vision-language embedding. This is then channelled through two heads:
\vspace{-0.08cm}
\begin{enumerate}[leftmargin=*]
    \item The task planning head takes the encoded VL embedding through an autoregressive LSTM module consisting of two layers. This component encodes historical context of the  previous predicted outputs, which is then passed through two linear classifiers to predict the  \textit{(action, object)} pairs associated with the input language and image features.
    
    \item The mask prediction head uses the current RGB first party view, the encoded language and the action-object prediction hidden state from the task planning head to predict a single mask referring to the object to be interacted with. We use the referential image segmentation architecture described in \cite{wang2021cris} for predicting a single mask for the interactable object instance that the language instruction is referring to.
\end{enumerate}
The above procedure is repeated for a natural language instruction and a series of first party views until an End-of-Sequence (EOS) flag is emitted by the task planning head (which is modelled as a separate binary classifier for every time step). 

For training, we use a learning rate of $1e^{-6}$ for the backbone and mask prediction head, and $1e^{-4}$ for the action prediction head. We use the Adam optimizer with a weight decay of $1e^{-5}$ and a learning rate decay of $0.1$ after 35 epochs. The model is trained for 65 epochs with a global batch size of 64 (16 per GPU) on 4 Tesla V100 GPUs .

\section{Results and Discussions}
We present the results on the trajectory dataset. The models are trained on the training split, and evaluated on the validation split. 

\subsection{Evaluation Metrics}

\subsubsection{Agent Evaluation}
We expect that a good embodied agent should be able to finish the missions efficiently. To achieve this goal, the agent should understand the human instructions and generate a corresponding sequence of actions. Thus, we evaluate the agent using the following metrics:

\begin{itemize}[leftmargin=*]
    \item \textbf{Mission success rate (MSR).} For each mission, there is a mission success variable $m$ indicating whether the goal conditions have been met for the mission. $m=1$ if all the goal conditions are met. Mission completion rate is calculated by averaging $m$ across all the missions. 
    \item \textbf{Average number of robot actions (NRA).} To measure the efficiency of task completion, we also record the number of actions taken by the agent to complete each mission. Average number of robot actions is calculated by averaging the number of actions across all the missions. 
\end{itemize}

\subsubsection{Vision Model Evaluation}
The Neural-Symbolic model is composed of a separate visual component that performs instance level segmentation of the first party view and/or surrounding images of the agent. We evaluate the instance segmentation model using the standard COCO evaluation metric, the Mean Average Precision (mAP). The mAP metric is calculated by averaging the precision at Intersection over Union (IoU) thresholds ranging from 0.5 to 0.95 in steps of 0.05. The score of the predicted instance class is not taken into consideration, and instead a cap is applied on the number of maximum detections per image (chosen to be a 100). To be able to choose an operating point for our end-to-end model and also tune for recall, in addition to the standard COCO mAP metric, we also provide a score and iou thresholded precision metric, which we call \textit{t-mAP}. This is calculated by thresholding the output instance class scores at [0.05, 0.1, 0.3, 0.5, 0.7],
and thresholding the IoUs at [0.1, 0.3, 0.4, 0.5, 0.75, 0.8]
for all 30 (score, IoU) combinations and then averaging them. We present both metrics for all objects, as well as metrics categorized for small, medium and large objects. The results are provided in \Cref{tab:vision_eval}.

\subsection{Results and Analysis}
\begin{table*}[t]
\begin{subtable}{\textwidth}
    \centering
    \begin{tabular}{||l|ccc||}
    \hline
    \textbf{Category} & \textbf{Area ($px^{2}$)} &\textbf{COCO mAP} & \textbf{t-mAP}  \\ \hline
    Small & 0 - 1296 & 37.63 & 91.49  \\ 
    Medium & 1296 - 9216 &60.41 & 92.15  \\  
    Large & 9216 - 90000 & 64.72 &84.91 \\  
    \hline
    Overall & 0 - 90000 & 46.03  &  89.5  \\ 
    \hline
    \end{tabular}
    \vspace*{0.1 cm}
    \caption{Image segmentation results for the vision model used in the neural-symbolic approach. We present the metrics for small, medium and large objects.}
    \vspace*{0.3 cm}
    \label{tab:vision_eval}
\end{subtable}
\begin{subtable}{\textwidth}
    \centering
    \begin{tabular}{||l|cc||}
    \hline
    \textbf{Method} & \textbf{MSR (\%)} & \textbf{NRA}  \\ \hline
    Neural-Symbolic & 18.19 & 11.82 \\ 
    VL-model & 22.80 & 12.73 \\ 
  Neural-Symbolic (with QA) & 19.32 & 11.73 \\ 
    VL-model (with QA) & \textbf{34.20}  &  \textbf{8.82} \\
    \hline
    \end{tabular}
    \vspace*{0.1 cm}
    \caption{Experimental results for agent evaluation on validation dataset for both baseline methods. MSR stands for mission success rate. NRA stands for average number of robot actions. }
    \vspace*{0.3 cm}
    \label{tab:result}
\end{subtable}
\begin{subtable}{\textwidth}
    \centering
    \begin{tabular}{||l|cccc||}
    \hline
     \textbf{Mission} & \textbf{NS (\%)} & \textbf{NS (\%)} & \textbf{VL (\%)} & \textbf{VL (\%)} 
     \\
     \textbf{Type} & w/o QA & w/ QA & w/o QA & w/ QA
     \\ \hline
     breakObject & 0.00 & 0.00 & 21.11 & 41.11  \\ 
     clean\&deliver & 12.64 & 13.79 & 13.79 & 19.10 \\ 
     color\&deliver & 0.00  & 0.00 &  0.00 & 0.00 \\ 
     fill\&deliver & 14.58  & 18.75 & 10.41 & 22.91 \\ 
     freeze\&deliver & 33.33  & 25.00 & 0.00 & 8.33 \\ 
     heat\&deliver & 5.13 & 5.13 &  10.25 & 28.20\\ 
     insertInDevice & 14.12  & 14.69 & 14.68 & 20.90 \\
     pickup\&deliver &  9.47  & 12.63 & 15.43 & 27.36 \\ 
     pourContainer & 14.53  & 16.24 & 16.23 & 30.76 \\ 
     repair\&deliver & 11.11  & 12.96 & 9.25 & 29.62 \\ 
     scanObject & 41.44  & 41.44 & 37.83 &  56.75 \\ 
     toggleDevice & 57.14   & 56.19 & 81.90 & 81.90 \\ \hline      
    \end{tabular}
    \vspace{0.1 cm}
    \caption{MSR of the Neural-Symbolic (NS) and Vision Language (VL) models based on mission type. The results are for with QA and without QA.}
    \label{tab:mission_sr}
\end{subtable}
\vspace{-0.8 cm}
\caption{Baseline Model Evaluation Results}
\end{table*}

The overall mission completion results from the baseline models are displayed in \Cref{tab:result} and \Cref{tab:mission_sr}. Both models are evaluated with a cap of 50 maximum allowed steps per mission, and a maximum of 10 failed steps per mission, beyond which if the mission goal is not completed, the agent stops execution leaving the goal incomplete. Below, we analyze the results for both the models and provide insights into the model performances.

\subsubsection{Neural-Symbolic Model}
\begin{figure}[t]
    \centering
    \includegraphics[width=0.3\textwidth]{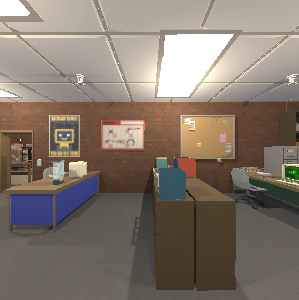}
    \caption{There are two tables in this image, a blue table on the left and a green one on the right, leading to challenging disambiguation tasks for the agent if not guided adequately.}
    \label{fig:multi_instance}
\end{figure}

Overall, adding QA leads to slight improvement of performance for the Neural-Symbolic model. Looking at the MSR for each type of mission (\Cref{tab:mission_sr}), we notice that the model performs well for missions with short horizons (e.g. scanObject, toggleDevice). As an ablation study, we also evaluate the multi-modal transformer on the validation set by giving it the ground-truth visual observations at each time step. As a result, the model can correctly predict all the actions and objects, including the rotation actions, for 71.3\% of the missions without questions and answers, and 76.9\% with questions and answers. These quantitative results demonstrate that the bottleneck for the neural-symbolic model is not the action and object prediction. In fact, since it is common that there are multiple objects of the same class in a room, even if the transformer model correctly predicts the object class and the vision model correctly predicts all the object masks of that class in the image, since we use heuristics to do visual grounding, the model can still make mistakes selecting the mask for the right object instance (\Cref{fig:multi_instance}). This can also explain why the improvement with QA is limited, since the language instructions do not participate in the grounding once the action and object are predicted. In addition, the vision model has difficulties detecting some small objects, such as carrot, coffee cup, hammer and screw driver, which also causes some errors. Some objects look similar but belong to different classes, such as the time machine and the microwave, and the table and book shelf, making it hard for the vision model to distinguish.  

\subsubsection{Vision Language Model}
The VL model shows a 11.40\% absolute improvement in MSR when trained and evaluated with QA compared to training and evaluating without QA. The agent is also able to complete the missions more efficiently with QA, by showing an improvement of 3.91 in average number of steps for mission completion. This demonstrates that asking the right questions improves task execution efficiency and accuracy. 
The agent shows significant improvement in MSR for task types for which the QA adds valuable information. For tasks like \textit{toggling} which few objects can afford, the instructions tend to already be descriptive enough in natural language (\textit{"Turn on the red computer", "Press the blue button"}), which explains why adding QA shows no improvement. For all other task types which involve numerous objects, the QAs provide crucial additional information that is not naturally provided with the instruction, thus leading to significant performance improvement. From analysis, the \textit{color \& deliver} missions are prone to failure because of difficulty in disambiguating the receptacle of \textit{delivery}.

While the VL model demonstrates improved performance using an end-to-end method that generates actions and masks simultaneously, we identify key insights, sources of errors and failure points of the end-to-end architecture by analyzing the robot trajectory for 25 missions: 
\begin{enumerate}[leftmargin=*]
\item Some missions fail because of failure to visually ground the natural language instruction to a single object instance for interaction in the case of multiple instance candidates. E.g. in \Cref{fig:multi_instance}, there are two \textit{desks} - blue on the left and green on the right. If the natural language instruction is not descriptive enough, the model fails to visually ground the \text{desk} of interest. On analysis, we notice that in the \textit{<x>\&deliver} mission type, the agent is often able to complete the \textit{<x>} goal and fails at the \textit{deliver} goal because of ambiguity in language (\textit{"The desk is right in front of you with the monitor on it"}), or model visual perception failures to disambiguate between multiple instances of the same object. 

\item Sometimes, if the agent veers off its path due to a wrong prediction (e.g. if it goes to a wrong room), the subsequent instructions to find an object in the room steers the agent down a path to infinitely rotate about its position to try finding the object ,i.e., the agent is trying to respond to the next instruction that assumes successful attainment of the correct previous state. This can potentially be solved by better error correction mechanisms in the model to identify such loopy states and exit them to explore other options in the trajectory, or to execute a strategy to backtrack to previous instructions. Since the current baseline models use the readily available answers in the dataset, this can also be improved in models that are designed to ask questions based on its 
real-time current visual observations.

\item Since this is an end-to-end model, the language and visual features are jointly encoded and trained end-to-end from input language and images to output masks and actions. Some errors arise because the model is attending more to the visuals in front of it rather than the language input. E.g. if the robot is standing very close to a cabinet door, or a shelf with a coffee pot, even though the language instruction was to \textit{"go to the laser cannon"}, it will try to  \textit{Open} and  \textit{Close} the cabinet door, or \textit{Pickup} the coffee pot, respectively, which the model has learnt from the data.

\item Some more sources of error distributed in the model across all evaluation tasks include ambiguity in language, inability to model long range task instructions like \textit{"turn right, go to the \textit{breakroom}, heat the bowl and place it on the table"} - which consists of at least 9 steps, inability to detect masks for some very small objects, or for objects too up close, and failure to distinguish between very similar looking objects (like \textit{Laser Cannon} and \textit{Freeze Ray}). These are common issues between the Neural-Symbolic and the VL models. 

\item We also did an ablation study by training only the mask prediction head and using the ground truth action predictions. Evaluating this model on the validation dataset yields a MSR of 71.89\%. This simulates the scenario when the perfect actions are predicted to situate the robot in a position to predict a mask for an interactable object, thus evaluating only object detection and visual grounding of the natural language instruction. This result shows the model's difficulties in finding object locations in an end-to-end fashion, requiring the incorporation of better path planning modules that learn from multi-modal inputs. 
\end{enumerate}

\section{Conclusion and Future Work}
In this paper, we introduce Alexa Arena, a user-centric Embodied AI platform. Alexa Arena features user-friendly graphics, animations and control mechanisms to facilitate user-centric EAI research at scale. Presented as games to general users, missions in Alexa Arena involve a variety of meticulously designed object categories and interaction actions, opening up new possibilities for data collection, EAI development and evaluation. We present one use case of the platform on dialogue-guided task completion, and also release a hybrid dataset with challenging problems in task planning, visual grounding and natural language understanding. We provide baseline results in a dialog augmented setting for two model architectures. In future work, we aim to incorporate a model-based questioning mechanism to which an oracle provides relevant answers during task execution, leading to more accurate and dynamic models that use real-time contextual information for grounding. Another direction is to enable the human and the robot to control different embodied agents and create more scenarios for HRI. We also plan to release more missions and scenes in future updates to incorporate other exciting elements in Arena, like gravity flippers, hazards and other blocker objects in the agent's execution path, opening up avenues for innovation in multi-modal reasoning and compositional task planning.

\bibliography{main}
\bibliographystyle{icml2021}



\pagebreak

\appendix

\section{Appendix}
\label{sec:appendix}

\subsection{Data Collection}

\begin{figure*}[t]
\centering
    \begin{subfigure}[b]{0.85\textwidth}
    \flushleft
    \includegraphics[width=\textwidth]{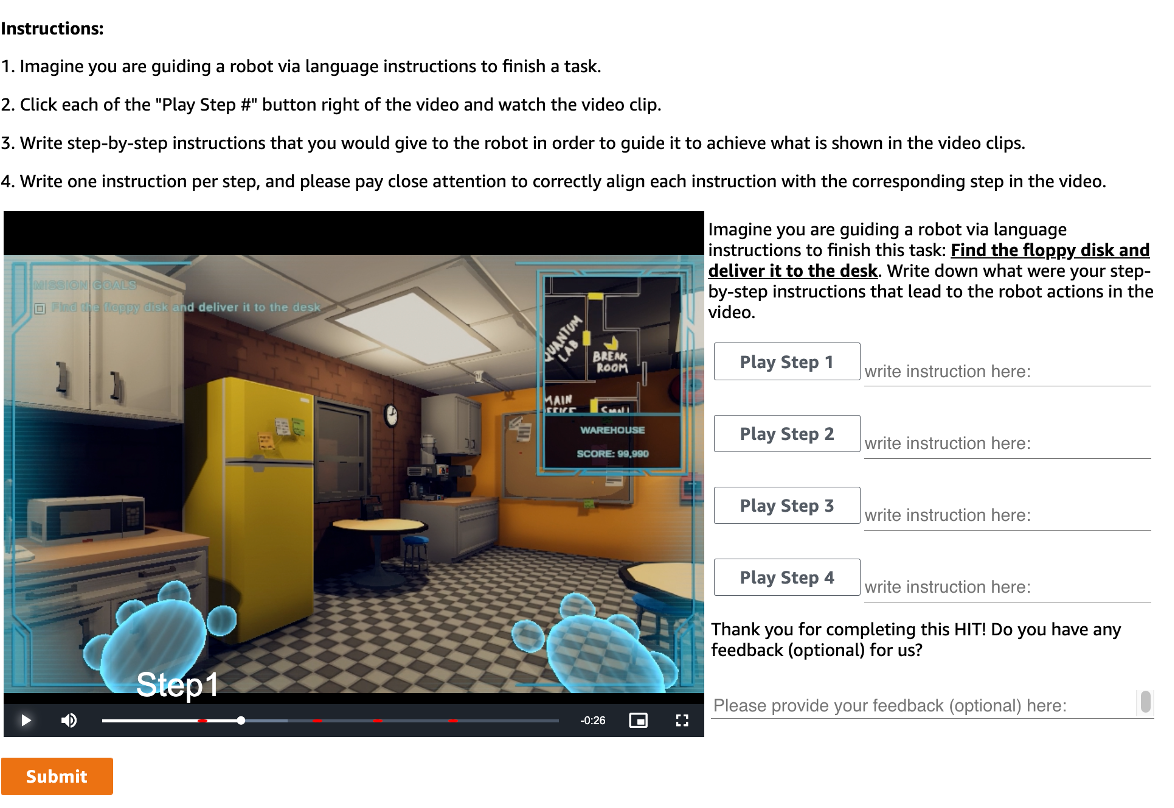}
    \caption{User interface for collecting human instructions. Annotators watch the expert demonstrations and write down instructions to guide the robot.}
    \label{fig:ui_instruction}
    \end{subfigure}

    \begin{subfigure}[b]{0.85\textwidth}
    \flushleft
    \includegraphics[width=\textwidth]{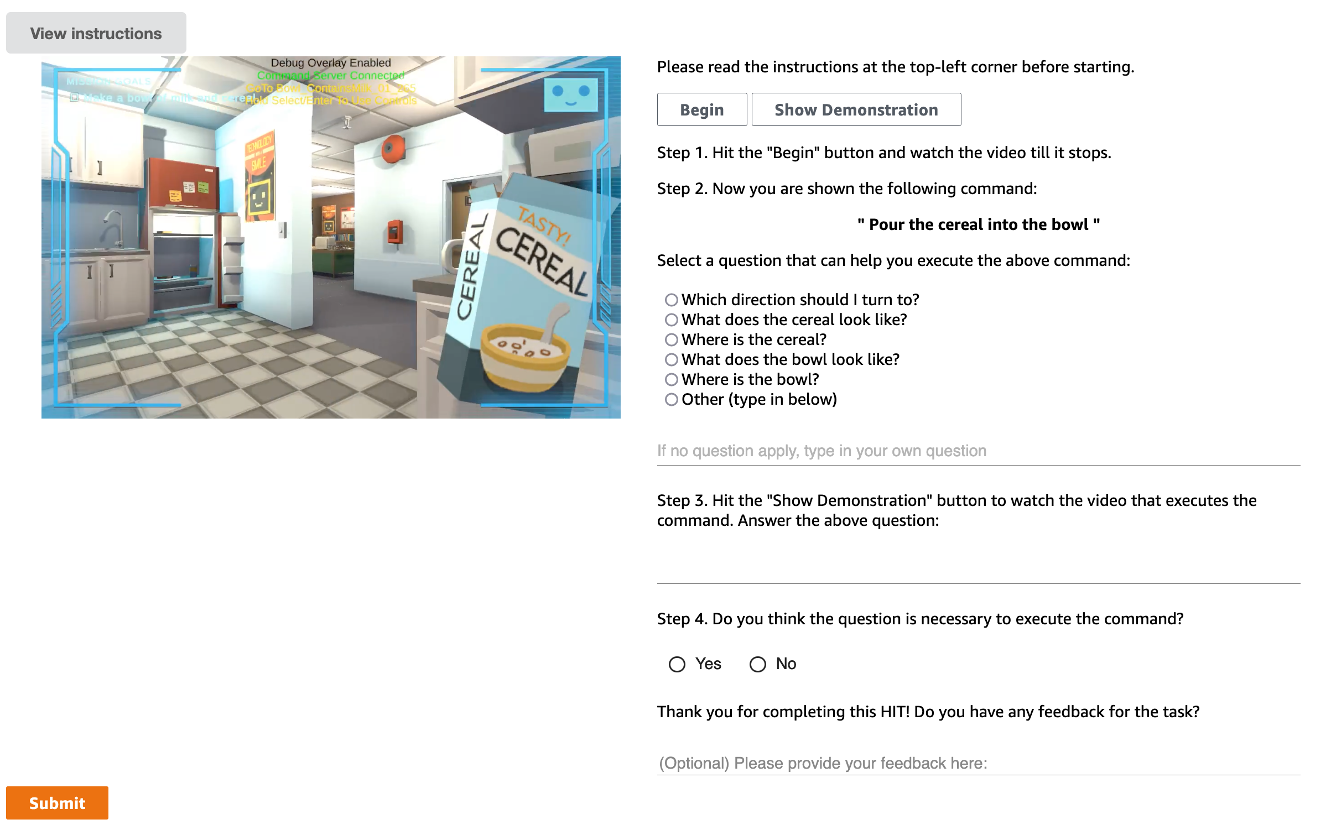}
    \caption{User interface for collecting questions and answers. Annotators input their questions and answers to help the robot complete the mission.}
    \label{fig:ui_qa}
     \end{subfigure}
     \caption{User interface for data collection.}
\end{figure*}

\subsubsection{Human Language Annotation}
\label{appendix:annotation}

To collect natural language dialogue for the missions, we design a two-stage data annotation process on Amazon Mechanical Turk (AMT). The first stage is for collecting instructions and the second is for questions and answers. In both stages, the annotator watches a video containing an expert demonstration for the mission and provide the annotations by text data or multiple choice questions.

In the first stage, annotators are told to write instructions to tell a “smart robot” how to accomplish a task. During the process, an annotator first watches the video of the ground-truth robot actions, then writes instructions for each highlighted video segments. \cref{fig:ui_instruction} shows the user interface for collecting language instructions. 

After all the instructions are collected, we start the second stage of annotation, where the annotators are asked to raise questions to help them complete the task, as if they are controlling the robot to follow the instructions. They also need to answer their own questions. \cref{fig:ui_qa} shows the interface for collecting questions and answers. Here are the steps annotators need to follow:
\begin{enumerate}[leftmargin=*]
    \item Annotator watches a video: A 10 second video clip is played till the beginning of a sub-task, and the respective instruction is shown on screen. The video clip helps the human annotator to understand the initial environment states.
    \item Annotator selects a question: The annotator selects one pertinent question that they think may help task completion. The questions are generated from several predefined templates. Annotators also have the options to type in their own questions.
    \item Annotator answers the question: The same annotator watches the video segment of an expert agent working on the sub-task specified by the language instruction, and then providing the answer to the question in text. 
    \item Annotator indicates whether asking the question is necessary: In some cases, the instruction and the visual context is already clear enough, and asking a question is not necessary.
\end{enumerate}




To ensure the annotation quality, in practice, we collected 3000 data sessions for the first data collection stage. Then we asked additional annotators to identify annotation errors and try to correct them, e.g. erroneous language instructions, grammatical errors, misalignment between instructions with robot actions, etc. For data sessions where most collected annotations do not make sense and are not worth fixing, the additional annotators are asked to simply ignore them. As a result, 20.3\% sessions are ignored and 35.7\% instructions are corrected. From the data verification results, we identified a subset of AMT workers (around 300 workers) who generated high-quality data, and used them for our subsequent data collection process.

\subsubsection{Questions Generation}

Similar to \cite{gao2022dialfred}, the questions choices in the second stage of annotation are generated using predefined templates. Given an instruction, we extract nouns and insert them into the templates to construct questions. The nouns extracted correspond to query objects that the agent need more information to complete the mission.

In particular, we consider three types of questions, related to the location and appearance of the query object $o$, and the relative direction between the agent and the target position. The templates for each question are defined below:
\begin{enumerate}[leftmargin=*]
    \item Location: where is $o$?
    \item Appearance: what does $o$ look like?
    \item Direction: which direction should I turn to?
    \item Reference: which $o$ are you referring to?
\end{enumerate}

\begin{figure*}[t]
    \centering
     \begin{subfigure}[b]{0.32\textwidth}
         \centering
         \includegraphics[width=\textwidth]{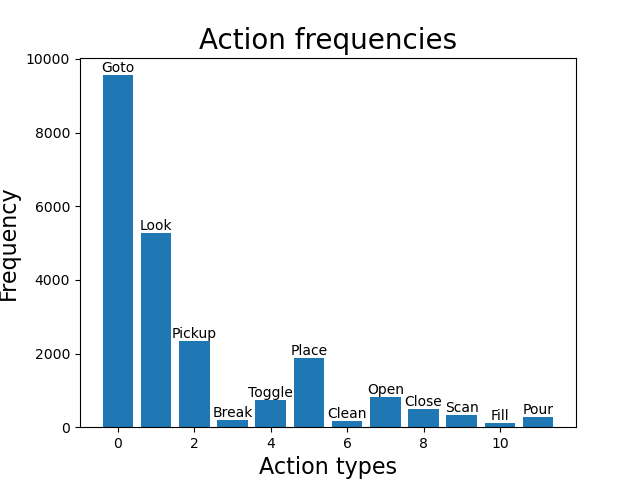}
         \caption{Frequencies of action types.}
         \label{fig:action_freq_train}
     \end{subfigure}
     \begin{subfigure}[b]{0.32\textwidth}
         \flushleft
         \includegraphics[width=\textwidth]{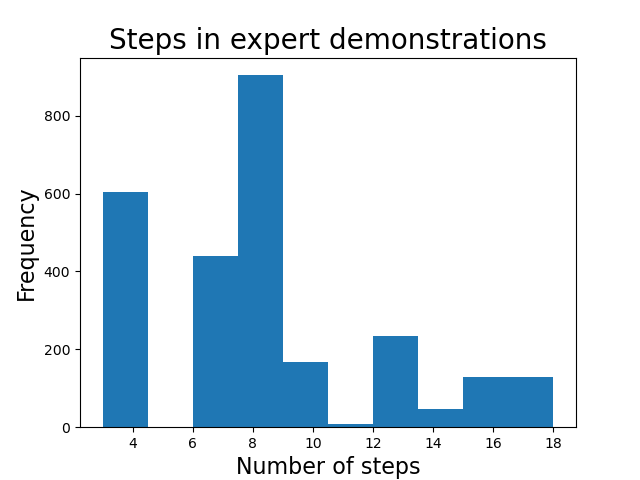}
         \caption{Steps in expert demonstrations.}
         \label{fig:expert_steps_train}
     \end{subfigure}
     \begin{subfigure}[b]{0.32\textwidth}
         \flushleft
         \includegraphics[width=\textwidth]{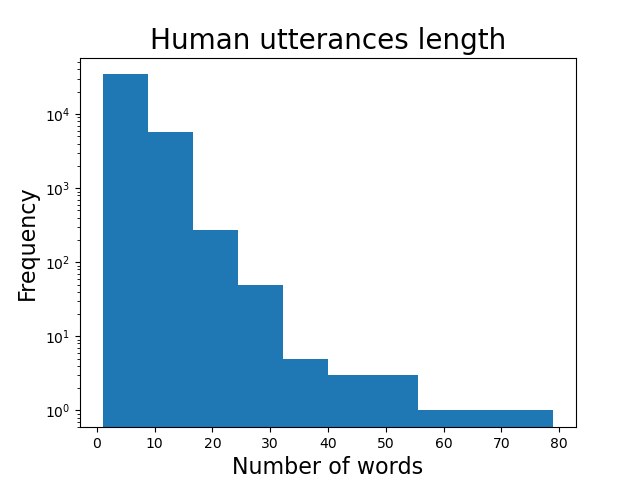}
         \caption{Length of human utterances.}
         \label{fig:human_utt_length_train}
     \end{subfigure}
    \caption{Statistics of the \textit{training} split of the trajectory dataset.}
    \label{fig:train_data_stats}
\end{figure*}

\begin{figure*}[t]
    \centering
     \begin{subfigure}[b]{0.32\textwidth}
         \centering
         \includegraphics[width=\textwidth]{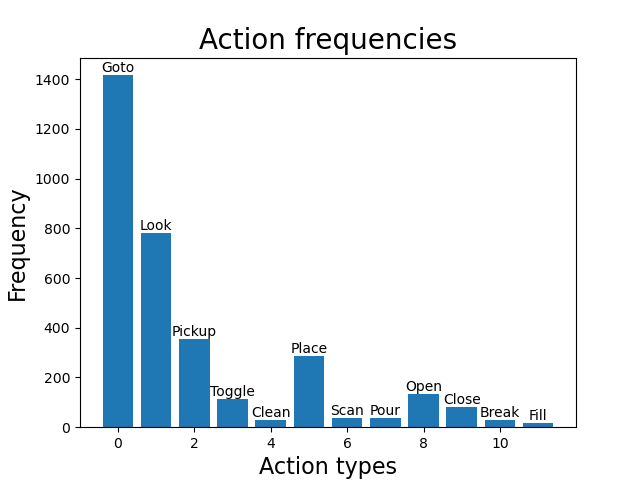}
         \caption{Frequencies of action types.}
         \label{fig:action_freq_valid}
     \end{subfigure}
     \begin{subfigure}[b]{0.32\textwidth}
         \flushleft
         \includegraphics[width=\textwidth]{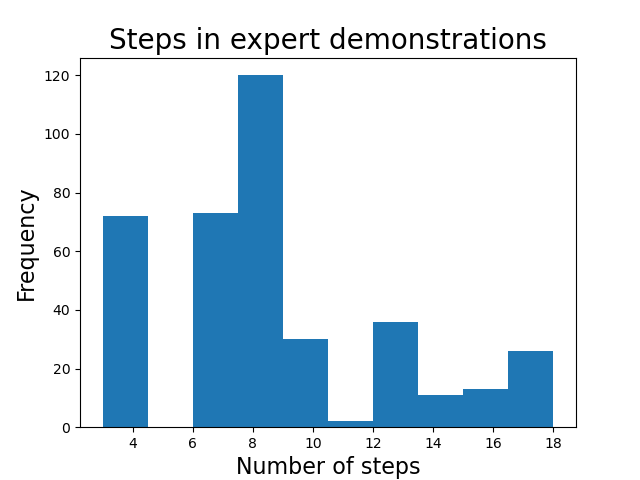}
         \caption{Steps in expert demonstrations.}
         \label{fig:expert_steps_valid}
     \end{subfigure}
     \begin{subfigure}[b]{0.32\textwidth}
         \flushleft
         \includegraphics[width=\textwidth]{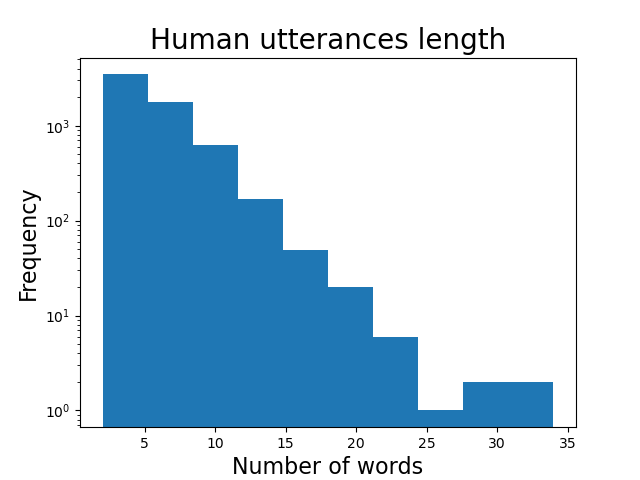}
         \caption{Length of human utterances.}
         \label{fig:human_utt_length_valid}
     \end{subfigure}
    \caption{Statistics of the \textit{validation} split of the trajectory dataset.}
    \label{fig:valid_data_stats}
\end{figure*}

\Cref{fig:train_data_stats} and \Cref{fig:valid_data_stats} shows the statistics of the trajectory dataset.

\subsubsection{Oracle Answers} 

Human language has a lot of variety and complexity, and thus can be quite challenging for the agent to understand. Alternatively, we can provide templated language to the agent as a starting point for language understanding. To this end, in addition to asking human annotators to answer the questions, we also build an oracle that can answer the questions by extracting information from the simulator metadata: (i) To answer the location question, we compute the direction of the object relative to the agent and the viewpoint that is closest to the object in the room. To help detect small objects, if the object is held by a container, we also provide the information of the container and other objects that are also held by the same container as landmarks; (ii) To answer the appearance question, we directly extract object shape, color and material information from the simulator metadata; (iii) For direction question, we compare the agent's location at the end of the sub-task with its initial location to provide answers for the agent's moving direction. After we have the necessary information, we use language templates to generate the answers. Example templates include: 
\begin{enumerate}[leftmargin=*]
    \item Location: The $o$ is to your [\textit{direction}] in/on the [\textit{container}] next to the [\textit{landmark}] in the [\textit{room}]. It is closest to [\textit{viewpoint}].
    \item Appearance: The $o$ is [\textit{shape}] and of [\textit{color}]. It is made of [\textit{material}].
    \item Direction: You should turn [\textit{direction}] / You don't need to move.
\end{enumerate}

\begin{table}
\begin{subtable}{0.48\textwidth}
    \centering
    \begin{tabular}{||l|c|c||}
    \hline
     \textbf{Mission Type} & \textbf{Train} & \textbf{Valid}  \\ \hline
     breakObject & 196 & 30 \\ 
     clean\&deliver & 161 & 29 \\ 
     color\&deliver & 29 &  4 \\ 
     fill\&deliver & 108 & 16 \\ 
     freeze\&deliver & 76  & 8 \\ 
     heat\&deliver & 42 &  13 \\ 
     insertInDevice & 381 &  59 \\
     pickup\&deliver &  669 &  95 \\ 
     pour & 265 & 39 \\ 
     repair\&deliver & 131 & 18 \\ 
     scanObject & 324 &  37 \\ 
     toggleDevice & 279  & 35 \\ \hline      
    \end{tabular}
    \vspace*{0.1 cm}
    \caption{Distributions of mission types across training and validation splits of the trajectory dataset. }
    \vspace*{0.3 cm}
    \label{tab:mission_types_distr}
\end{subtable}
\end{table}

\begin{figure*}[t]
    \centering
    \begin{subfigure}[b]{0.4\textwidth}
         \flushleft
         \includegraphics[width=\textwidth]{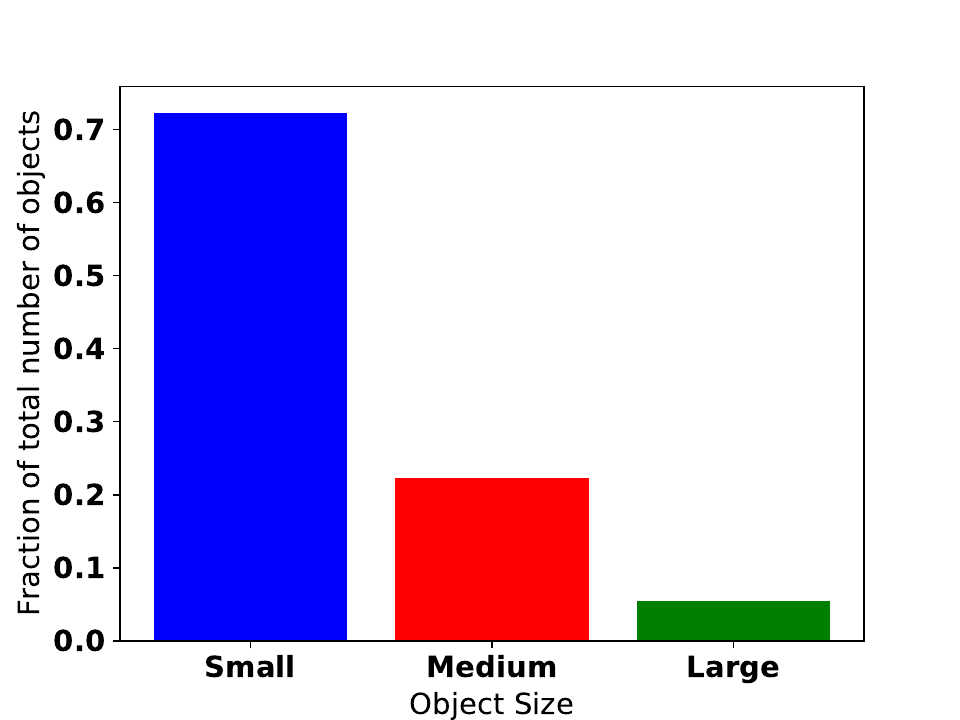}
         \caption{Vision data object size distribution.}
         \vspace{0.4cm}
         \label{fig:vis_data_area_stats}
     \end{subfigure}
     \begin{subfigure}[b]{0.9\textwidth}
         \flushleft
         \includegraphics[width=\textwidth]{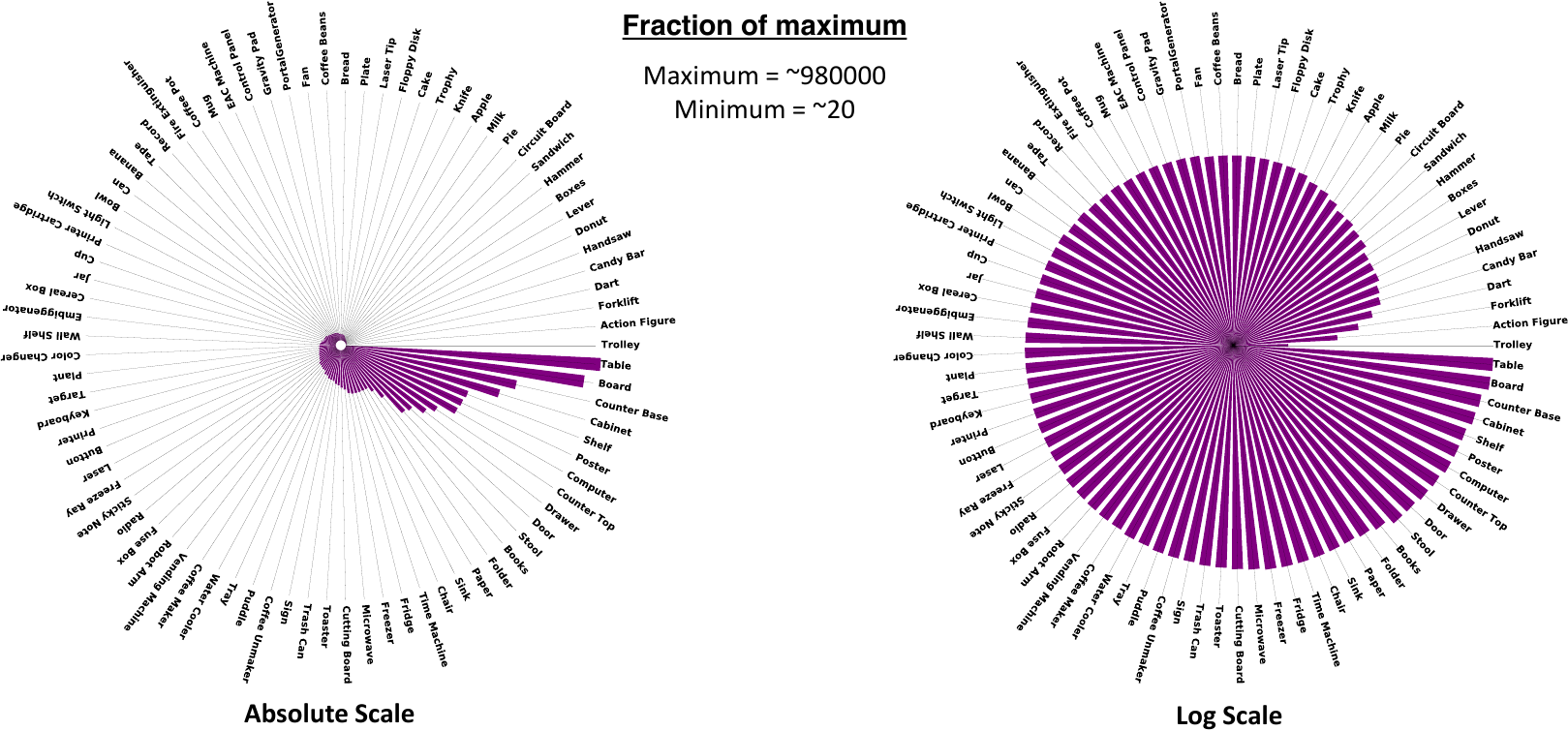}
         \caption{Vision data class frequency.}
         \label{fig:vis_data_stats}
     \end{subfigure}
    \caption{Vision dataset statistics}
\end{figure*}

\begin{figure*}[t]
    \centering
    \begin{subfigure}[b]{0.47\textwidth}
         \centering
         \includegraphics[width=\textwidth]{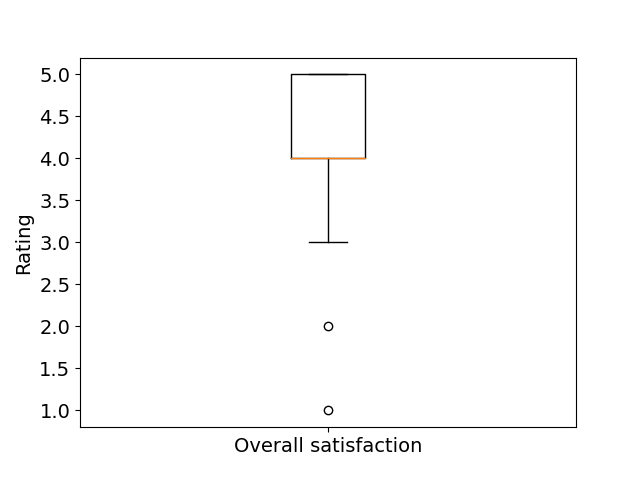}
         \caption{Overall satisfaction of the game, measured by a 5-point likert scale. 5 corresponds to "very satisfied" and 1 is "very dissatisfied". The orange line shows the median value.}
         \label{fig:csat}
     \end{subfigure}
     \hfill
     \begin{subfigure}[b]{0.47\textwidth}
         \centering
         \includegraphics[width=\textwidth]{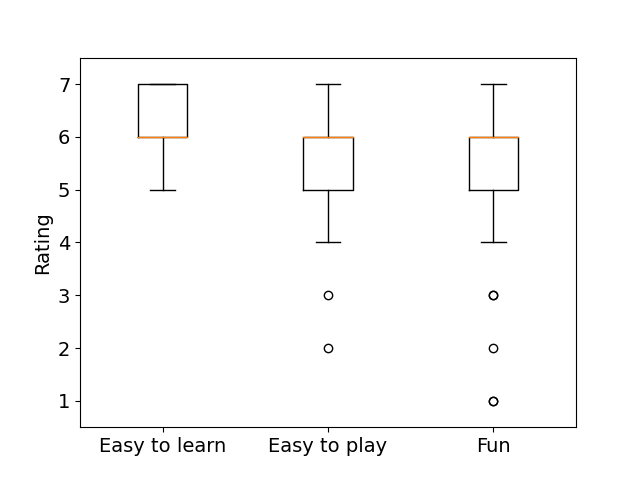}
         \caption{User perception towards the game, measured by a 7-point likert scale. 7 corresponds to "strongly agree" and 1 is "strongly disagree". The orange line shows the median value.}
         \label{fig:game_perception}
     \end{subfigure}
    \caption{User study results}
    \label{fig:user_study}
\end{figure*}

\begin{figure*}[t]
    \centering
    \includegraphics[width=0.5\textwidth]{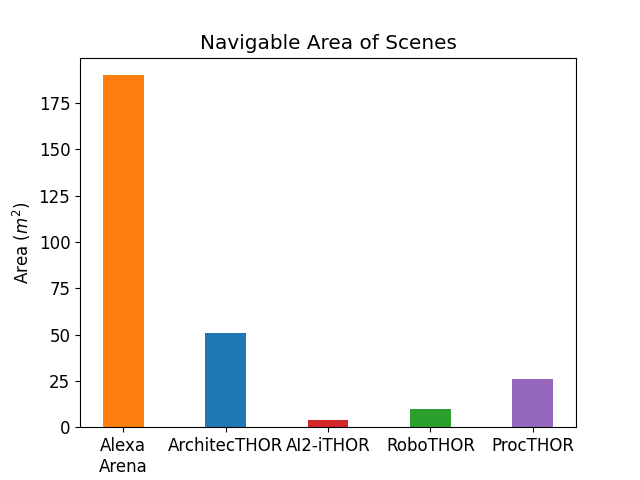}
    \caption{Bar plot of the navigable areas for Alexa Arena, comparing to ArchitecTHOR, AI2-iTHOR, RoboTHOR and ProcTHOR. }
    \label{fig:navigable_area}
\end{figure*}

\begin{figure*}[t]
    \centering
    \begin{subfigure}[b]{0.48\textwidth}
         \centering
         \includegraphics[width=\textwidth]{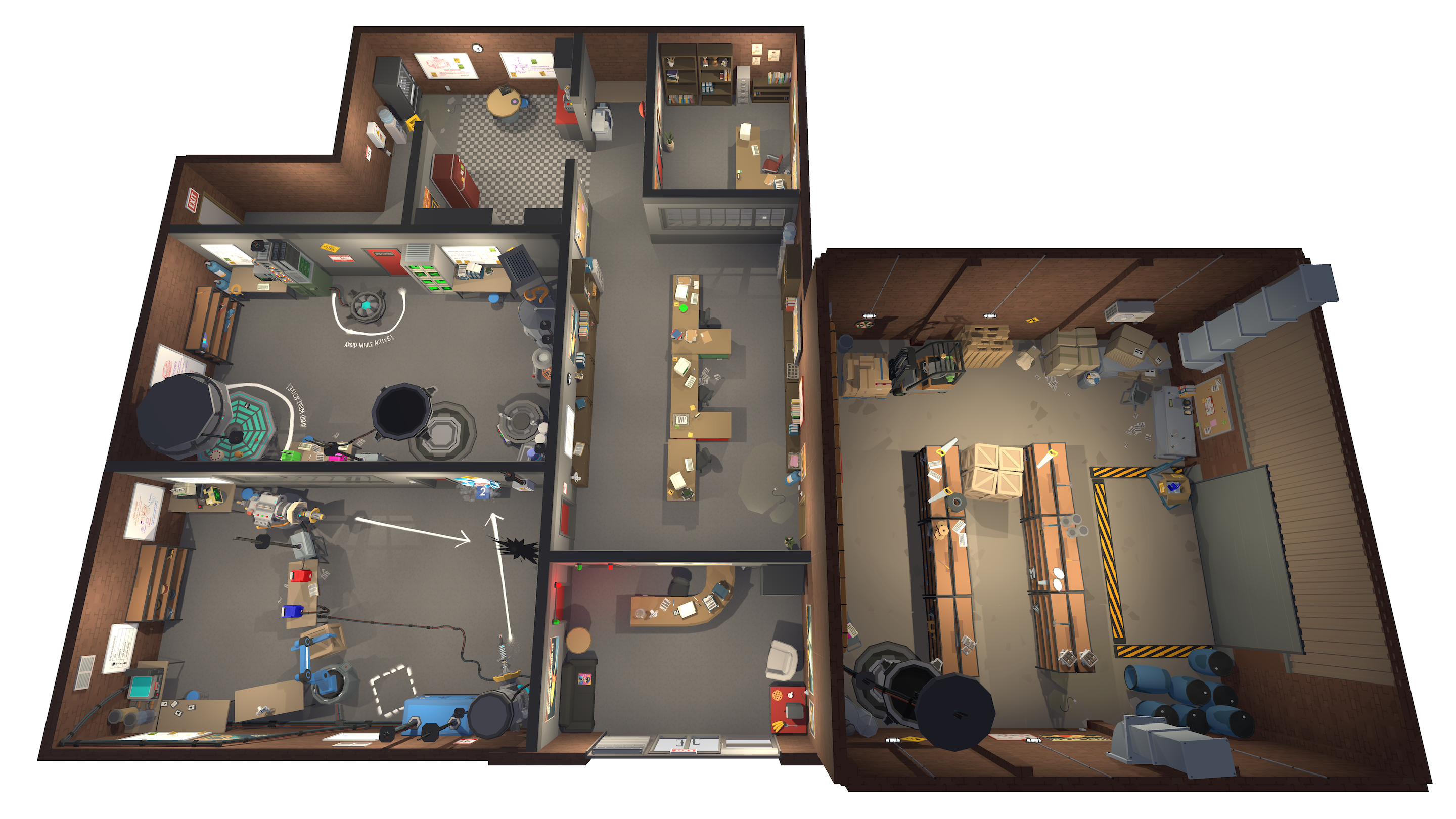}
         \caption{}
         \label{fig:layout1}
     \end{subfigure}
     \hfill
     \begin{subfigure}[b]{0.48\textwidth}
         \centering
         \includegraphics[width=\textwidth]{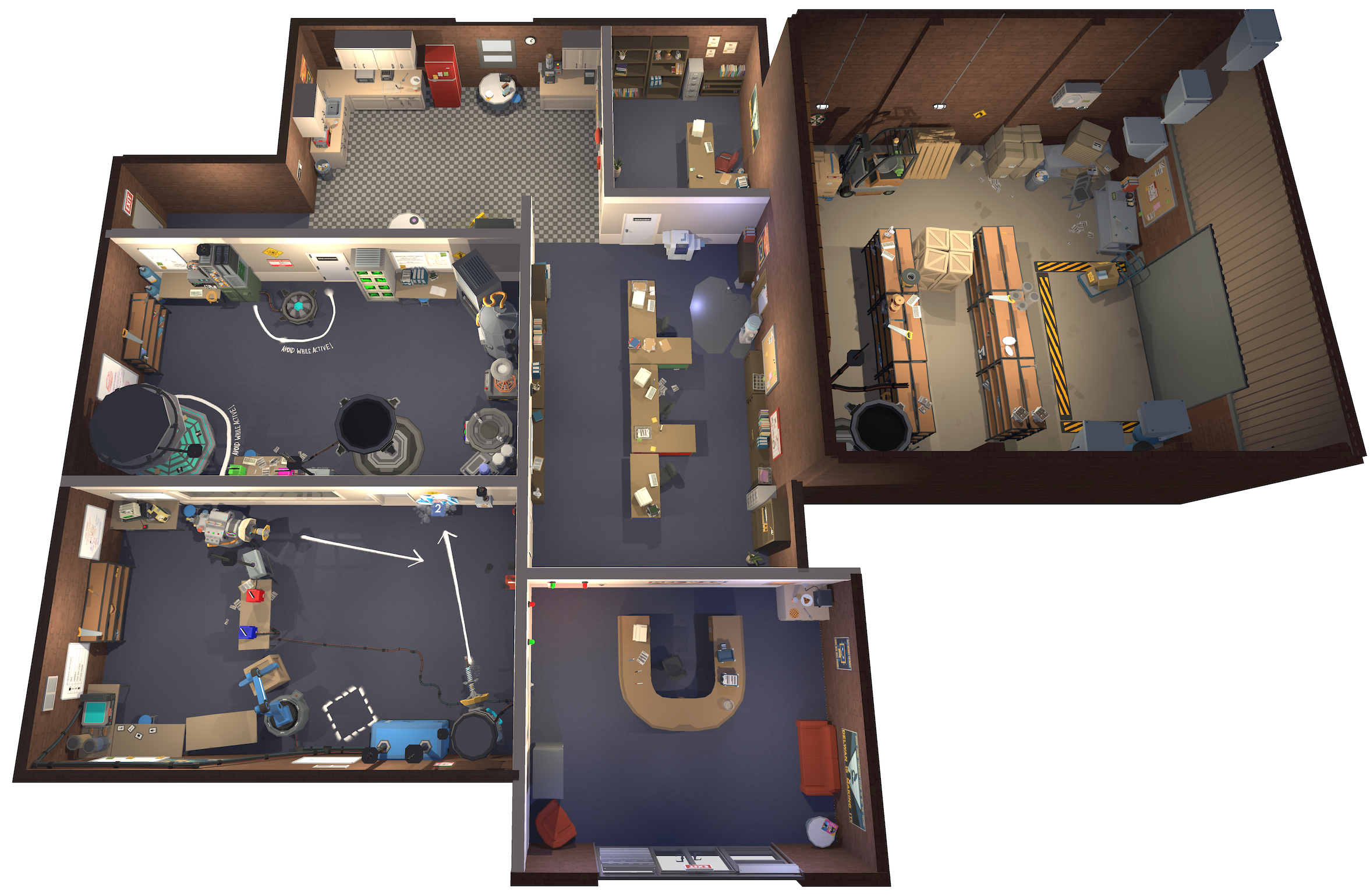}
         \caption{}
         \label{fig:layout2}
     \end{subfigure}
     
     \begin{subfigure}[b]{0.48\textwidth}
         \centering
         \includegraphics[width=\textwidth]{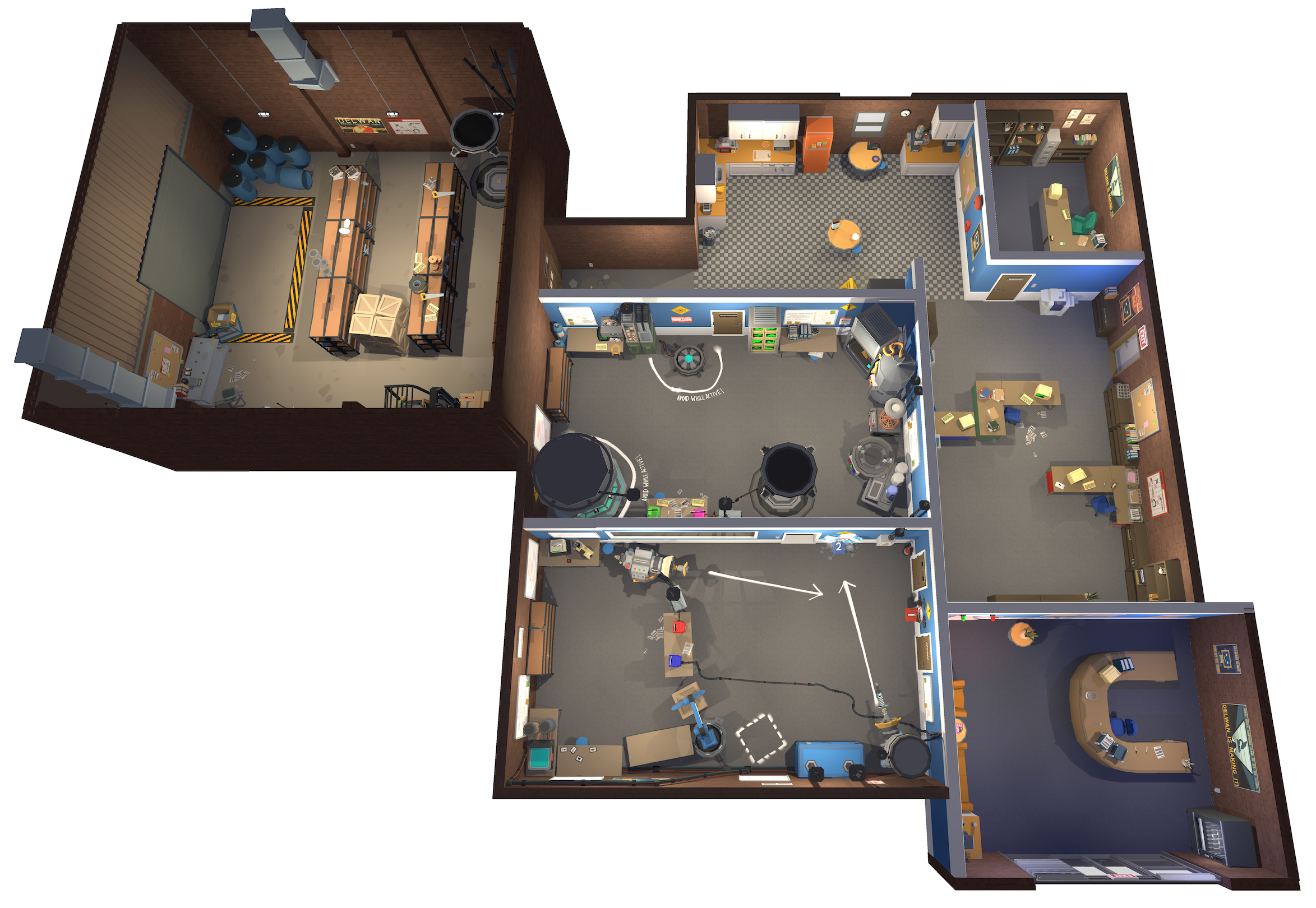}
         \caption{}
         \label{fig:layout3}
     \end{subfigure}
     \begin{subfigure}[b]{0.48\textwidth}
         \centering
         \includegraphics[width=\textwidth]{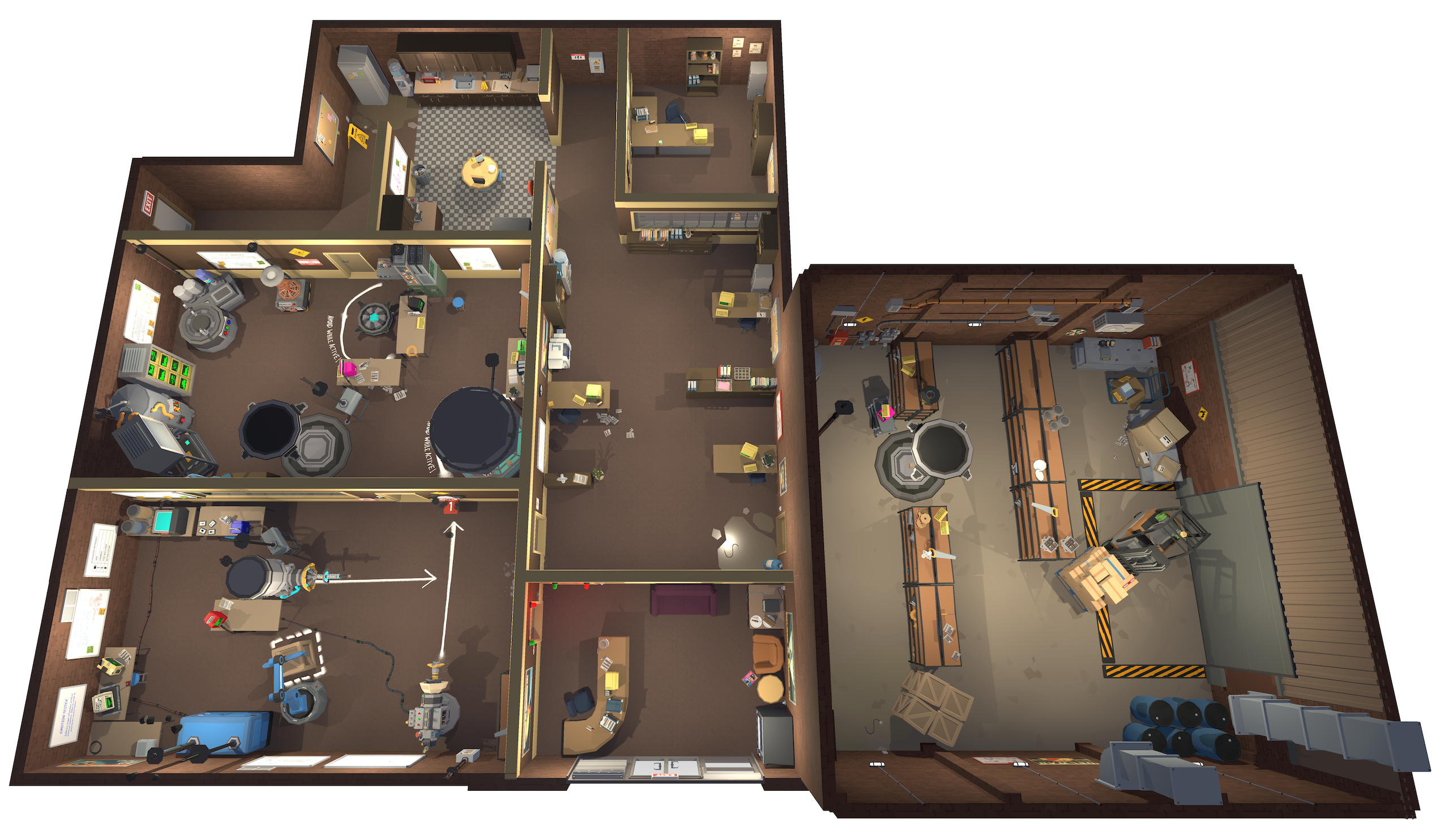}
         \caption{}
         \label{fig:layout4}
     \end{subfigure}
    \caption{Top-down images of 4 multi-room game-ready layouts in Alexa Arena.}
    \label{fig:layouts}
\end{figure*}


\subsubsection{Vision Data Generation}
\label{appendix:vision_data}

The image data collection strategy was simple and straightforward - the goal was to collect data spanning all possible objects spawnable in Arena from different distances, views and perspectives of the said object. This was done by following the below steps:
\begin{enumerate}[leftmargin=*]
\item Initialize a game mission using the CDF file with required objects. 
\item After initialization, we issue commands to navigate to all the different objects present in the scene using the primitive navigation actions and capture the first party view.
\item We then capture images by issuing interactive commands as per the affordances of that object at the time of capture, which are retrieved from the game metadata response.
\item We also capture additional perspectives by taking a random walk around the object using basic trigonometry.
\end{enumerate}

We repeat the above steps with various missions and objects with lights on and off for different lighting conditions. All the commands, object states and their locations in the scene, along with a host of other useful information are logged in metadata files. We provide the RGB color images, segmentation maps and associated metadata for each image. The statistics of the dataset are illustrated in \cref{fig:vis_data_stats}.

\textbf{Dataset Validation.} To validate the quality of the ground truth segmentation and bounding box annotations for modeling, we collect 1245 images containing instances spanning all object classes and annotate them with bounding boxes and associated class names. Since Arena features fantastical objects (like \textit{Gravity Flippers}), we also provide example images for each object class to familiarize the annotator with the appearance of all classes. The images are then validated by annotators according to the class names into one of the following 4 categories (with results provided in parenthesis):
\begin{enumerate}
    \item Yes, the object is fully contained within the bounding box (\textit{79.7\%})
    \item Yes, the object is partially contained within the bounding box (\textit{16.3\%})
    \item No, the object is not within the bounding box, but is in the image  (\textit{1.6\%})
    \item The object is not in the image  (\textit{2.4\%})
\end{enumerate}
For category 2, the objects are present only partially in the image because they are only partially visible in the robot's egocentric view at the time of image collection. Overall results show that the ground truth segmentation and bounding box annotations have very little noise and are of high quality to be used for training and evaluation of visual perception models. 

\subsection{User Study on Dialog-guided Task Completion}
To demonstrate the potential of Alexa Arena for real-time user interactions, we conduct a user study for dialog-guided task completion. The study is presented as games to the users. In the study, we run Alexa Arena on Amazon Echo devices. By talking to the devices, users can use verbal commands to control the agent in the virtual environment. 

\subsubsection{Experiment design}
\noindent \textbf{Participants.} 
We recruited 31 non-expert participants. 

\noindent \textbf{Metrics.} 
Overall, we look at the task performance and user satisfaction. We use the mission success rate for the task performance metric. For user satisfaction, we collect response on the overall satisfaction toward the game, as well as whether users believe the game is easy to learn, easy to play, fun to play and whether they would like to play it again. We use a 5-point likert scale for overall satisfaction, and 7-point likert scale for the other questions. 

\noindent \textbf{Tutorial.} 
Before working on tasks in the regular game, each participants is asked to complete a tutorial, which is designed to help them get familiar with the game mechanism. In the tutorial, the participant is asked to complete a tutorial mission, making a bowl of cereal, by giving verbal commands to the agent. For the first few required actions, such as picking up the cereal and pouring the cereal into the bowl, participants are given verbal instructions on the utterance they can use to execute the action. For the later actions, such as pouring the milk into the bowl, the participants are not given instructions on the utterance until they make mistakes. 

\noindent \textbf{Tasks.}
Each participant is asked to complete three regular missions: 
\begin{enumerate}[leftmargin=*]
    \item The first mission requires the agent to find the control panel and insert it into the laser. The agent also needs to turn on the red monitor that controls the laser to fire the laser canon. Note that there is also a blue monitor in the room which controls the freeze ray. It should not be used for this mission.
    \item The second mission requires the agent to pick up the bowl in the break room and place it on the color changer. After that, the agent needs to push the red button on the color changer to change the color of the bowl to red. Note that there are also green and blue buttons on the color changer, and the agent should not push them to complete this mission. 
    \item The third mission requires the agent to place a can of soda on a blue shelf which server as a target for the freeze ray. The robot can then toggle the blue monitor to fire the freeze ray and make ice cold soda.
\end{enumerate}

\noindent \textbf{Procedure.}
After entering the game, the user is asked to first complete the tutorial mission to get familiar with the game, before working on the regular missions. The order of the three regular missions are randomized to avoid order effect. The user can choose to skip the mission if they encounter any issues. During the game, users can check the sub-goals of the mission by looking at the text on the top left corner of the screen (\Cref{fig:web_tool}). The user can also read the sticky notes in the room to get some hints on how to complete the task. Sticky notes are displayed as flickering green dots in the minimap. After the interaction, the user is asked to complete a questionnaire for evaluation. 

\subsubsection{Results and analysis}
Out of the 31 participants, 15 participants are able to complete all three missions and 15 can complete two missions. Only 1 participant finishes one mission. This leads to an average mission success rate of 81.72\%. 

\Cref{fig:user_study} shows the results of the user study. In general, most users are satisfied with the game ($M=4.03, SD=0.93$) (\Cref{fig:csat}).  For game perception (\Cref{fig:game_perception}), most users agree that the game is easy to learn ($M=6.10, SD=0.73$), easy to play ($M=5.65, SD=1.23$) and fun to play ($M=5.26, SD=1.68$). As a result, 77.42\% of the participants would consider playing the game again. 

Looking at the free-form feedback, most participants like the game. We see comments like "It is first game playing on Alexa and I really loved the game.", "I thought the game was fun and figuring out game play was easy.", and "I think it has a good start. I like the idea of reaching some goals via voice commands." For improvement, users would like to see better language understanding, reduced game latency and more guidance in the tutorial. 

These results show that the Alexa Arena platform has the potential to reach general human users by creating missions as games, opening up the possibility for large-scale HRI data collection, system development and evaluation with humans in the loop. 

\section{Layouts}
Alexa Arena contains 10 large game-ready interative multi-room layouts. Each layout features an office-like environment with modular design, allowing the contents within rooms to be rearranged and repositioned. The layouts were hand designed to be visually different and aesthetically interesting. The office setting allows both mundane and fantastical interactive elements: users would be accustomed to seeing everyday objects in more conventional rooms (and would know instinctively how to interact with them), but entering rooms clearly labeled as "labs" they would expect to find futuristic devices and prototype machines that would require experimentation to discover their function. As a result of the above bipolar content, we are able to create game missions with imaginative situations sitting alongside normal everyday objects. The layouts in Arena are large: on average, each house has 190 square meters of navigable area, which is significantly larger than the scenes in other EAI platforms (\Cref{fig:navigable_area}). \Cref{fig:layouts} shows the top down images of 4 houses in Alexa Arena. 

\end{document}